\documentclass[pra,showpacs]{revtex4}
\usepackage{graphicx}
\usepackage{amsmath}
\usepackage{amssymb}
\usepackage{float}
\usepackage{algorithm}

\floatname{algorithm}{Protocol}
\usepackage{bm}

\usepackage{color}

\newtheorem{theorem}{Theorem}
\newtheorem{lemma}{Lemma}

\newtheorem{defn}{Definition}

\begin{document}
\title{Ground state blind quantum computation on AKLT state}  
\author{Tomoyuki Morimae}
\email{morimae@gmail.com}
\affiliation{Laboratoire d'Analyse et de Math\'ematiques Appliqu\'ees,
Universit\'e Paris-Est Marne-la-Vall\'ee, 77454 
Marne-la-Vall\'ee Cedex 2, France}

\author{Vedran Dunjko}
\affiliation{
School of Engineering and Physical Sciences,
Heriot-Watt University, Edinburgh EH14 4AS, U.K.
}

\author{Elham Kashefi}
\affiliation{
School of Informatics,
The University of Edinburgh, Edinburgh EH8 9AB, U.K.
}

\date{\today}
            
\begin{abstract}

The blind quantum computing protocols (BQC) enable a classical client with limited quantum technology to delegate a computation to the quantum server(s) in such a way that the privacy of the computation is preserved. Here we present a new scheme for BQC that uses the concept of the measurement based quantum computing with the novel resource state of Affleck-Kennedy-Lieb-Tasaki (AKLT) chains leading to more robust computation. AKLT states are physically motivated resource as they are gapped ground states of a physically natural Hamiltonian in condensed matter physics. Our BQC protocol can enjoy the advantages of AKLT resource states, such as the cooling preparation of the resource state, the energy-gap protection of the quantum computation, and the simple and efficient preparation of the resource state in linear optics with biphotons.

\end{abstract}
\maketitle  

\section{Introduction}
In a future, at the time when a scalable quantum computer is realized, 
there will be much demand for the blind quantum 
computation~\cite{Childs,Arrighi,blindcluster,Aha}, 
since the quantum computer must be super-expensive and super-fragile,
and therefore only limited number of servers can possess it.
In particular, it is likely that a client
(Alice) who has only a classical computer or
a primitive quantum instrument which is not sufficient for the universal
quantum computation
will ask a quantum server (Bob) to perform her quantum computation
on his fully-fledged quantum computer.
Bob runs his quantum computer according to
Alice's input and instructions,
and finally returns the final output of the quantum computation to Alice.
The crucial point in the blind quantum computation
is that Alice does not want Bob to learn her input, the algorithm she wants to run, 
and the final output of the computation.
For example, if Alice wants to factor a large integer by
using Shor's factoring algorithm~\cite{Shor},
the blind quantum computation must be performed in such a way that
Bob cannot know her input (the large integer), the output of the computation (a prime
factor), and even the fact that 
he is factoring. 

For the classical computation, Feigenbaum~\cite{Feigenbaum} introduced the notion of 
``computing with encrypted data",
and showed that for some functions $f$, an instance $x$ can be efficiently 
encrypted into $z=E_k(x)$ in such a way that Alice can recover $f(x)$
efficiently from $k$ and $f(z)$ computed by Bob. Moreover Abadi, Feigenbaum and Killian 
showed 
that no {\bf NP}-hard function~\cite{NP} 
can be computed blindly if unconditional security is required, unless the polynomial hierarchy collapses 
at the third level~\cite{Abadi}. Even restricting the security condition to be only computational \footnote{A crypto-system is unconditionally secure (also refereed to as ``information-theoretically secure") if it is secure even when the adversary has unlimited computing power. A weaker notion is computational security where the adversary power is restricted to efficient computation.}, the question of the possibility of blind computing, also known as fully homomorphic encryption, remined open for 30 years \cite{Gentry}.

Unlike classical computing, quantum mechanics could overcome the limitation of computational security. An example of the blind quantum computation was first proposed by Childs~\cite{Childs} where the quantum circuit model is adopted,
and the register state is encrypted with quantum
one-time pad scheme~\cite{onetimepad} so that Bob who performs
quantum gates learns nothing about information in the quantum register.
In this method, however, Alice needs to have a quantum memory and 
the ability to apply local Pauli operators at each step. Similarly the protocol proposed by Arrighi and Salvail \cite{Arrighi} requires multi-qubit preparations and measurements while Aharonov, Ben-Or and Eban's protocol \cite{Aha} requires a constant-sized quantum computer with memory.

On the other hand, in Broadbent, Fitzsimons and Kashefi's protocol \cite{blindcluster}, adapted to the one-way quantum computation~\cite{cluster,cluster2} all Alice needs 
is a classical computer and a very weak quantum instrument, which emits
random single-qubit states. In particular, she does not require any
quantum memory and the protocol is unconditionally secure. In their scheme, Alice's quantum instrument emits 
states 
$\frac{1}{\sqrt{2}}(|0\rangle+e^{i\xi}|1\rangle)$,
where
\begin{eqnarray*}
\xi\in {\mathcal A} = \Big\{\frac{k\pi}{4}\Big|k=0,...,7\Big\}
\end{eqnarray*}
is a random number which is
secret to Bob, 
and each state is directly sent to Bob.
Bob keeps all these states in his quantum memory,
and
creates the spacial resource state, which is alike to the two-dimensional
cluster
state and called 
``the brickwork 
state"~\cite{blindcluster}, by
applying the controlled-Z (CZ) operation among appropriate qubits
in the received states.
Alice calculates, by using her classical computer,
the direction in which a qubit should be measured,
and instructs the direction to Bob after some modifications such that
(i) the byproduct operators emerged from the previous Bob's measurements 
are compensated,
(ii) the random $\xi$ is canceled, and 
(iii) the true direction and the true measurement outcomes,
which reveal Alice's input, algorithm and output,
remain secret to Bob.
Bob does the measurement on the brickwork state 
according to the instruction from Alice,
and 
sends the measurement result to Alice for the next measurement.
After repeating such two-way classical communications between
Alice and Bob,
their measurement-based quantum computation is finally finished,
and Bob returns the final result (classical or quantum) of the computation to Alice.
As is shown in ~\cite{blindcluster},
Bob learns nothing about Alice's input, the algorithm she
wants to run, and the final output
of the computation.

In the recent active interaction between 
condensed matter physics and quantum information
science, plenty of novel resource states for the measurement-based
quantum computation beyond
the cluster state have been 
proposed~\cite{VerstraeteVBS,MiyakeAKLT,Gross,Gross2,Gross3,
Miyakeholographic,Miyake2dAKLT,
Wei,tricluster,Caimagnet,Li,reduction}.
These new resource states have several interesting features and 
advantages over the cluster
state. For example, 
some of those resource states are gapped ground states of 
their parent Hamiltonians, and
therefore they can be easily prepared by
cooling condensed matter systems
and the measurement-based quantum
computation can be protected from noise by the  
energy gap~\cite{MiyakeAKLT,Gross,Gross2,Gross3,
Miyakeholographic,Miyake2dAKLT,tricluster,Wei,Caimagnet,Li}.
In particular, Affleck-Kennedy-Lieb-Tasaki (AKLT) states~\cite{AKLT}
are physically motivated important
resource states~\cite{MiyakeAKLT,Gross,Gross2,Gross3,
Miyakeholographic,Miyake2dAKLT,Wei,Caimagnet,Li}, 
since they are gapped ground states of a physically
natural Hamiltonian which has long been studied in condensed matter physics,
and exhibit many novel features, such as
the one-dimensional Haldane phase~\cite{Haldane}, 
the diluted
antiferromagnetic order detected by the string order parameter~\cite{string},
and the effective spin-1/2 degree of freedom (edge state)
appearing on the boundary of the chain~\cite{edge}.
Furthermore, it was shown recently that
the preparation of the AKLT resource states is more efficient
and simpler than that of the cluster state in linear
optics with biphotons~\cite{Darmawan}.

It is, however, not obvious that these novel resource states also admit
blind quantum computation like the cluster state. They have 
drastically different properties compared to the cluster state or graph states in general,
such as the local purity,
the correlation property, and the entanglement property~\cite{MiyakeAKLT,Gross,Gross2,Gross3,Miyakeholographic,Miyake2dAKLT,Wei,tricluster,Caimagnet,Li}.
Furthermore, the actual computation over these new resources, such as the way of measurements and
the way of compensating byproducts, are also strongly different from those 
over the cluster state~\cite{MiyakeAKLT,Gross,Gross2,Gross3,Miyakeholographic,Miyake2dAKLT,Wei,tricluster,Caimagnet,Li}.

In this paper, we show for the first time that the ground state
measurement-based blind quantum
computation is possible with
AKLT resource states.
As we will see later, this is not a straightforward generalization
of the blind quantum computation on the cluster state,
although it borrows the general idea from \cite{blindcluster}. Hence a new proof of security for the blind quantum computation with AKLT states was also required.
As a result, our new protocol can enjoy the advantages of the AKLT resource states, namely
the easy preparation of the resource state 
by cooling condensed matter systems, 
the natural protection of the 
quantum computation by the energy gap,
and the simple and efficient
preparation of the resource state in linear optics with biphotons.

\begin{figure}[htbp]
\begin{center}
\includegraphics[width=0.5\textwidth]{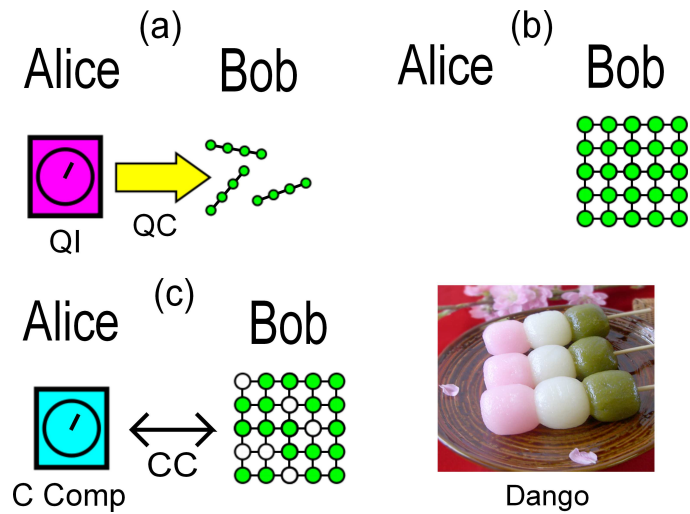}
\end{center}
\caption{(Color online.) 
The single-server protocol. 
(a): Alice's quantum instrument (QI) emits many four-qubit
states so-called ``Dango" states
each of which is directly sent to Bob through the quantum channel (QC).
Bob stores all of them in his quantum memory.
(b): From these Dango states,
Bob creates the resource state.
(c): Alice calculates the angle in which a particle
should be measured by using her classical computer (C Comp)
and sends the angle to Bob through the classical channel (CC).
Bob performs the measurement
and returns the result of the measurement to Alice.
They repeat this two-way classical communication until they finish the computation.
Bob finally sends the final output of the computation to Alice.
} 
\label{single}
\end{figure}

More precisely, we propose two methods, the single-server protocol
and the double-server protocol.
In the single-server protocol (Fig.~\ref{single}), 
there are two parties, called Alice and Bob.
Alice, the client, has a classical computer and
a quantum instrument which is not sufficient for the universal
quantum computation, whereas Bob, the server, has a universal quantum
computer.
Alice's quantum instrument emits random four-qubit states so-called
``Dango states"
each of which is directly sent to Bob through the quantum channel
(Fig.~\ref{single} (a)),
and Bob stores all of them in his quantum memory.
Each Dango state hides certain random number which is secret to Bob,
and these secret numbers are used later for the blind quantum computation.
From Dango states,
Bob creates the resource state, which is unitary equivalent
to an AKLT state (Fig.~\ref{single} (b)).
Alice calculates the angle in which a particle should be measured
by using her classical computer,
and sends the angle to Bob through the classical channel.
Bob performs the measurement according to Alice's information,
and returns the result of the measurement to Alice.
They repeat this two-way classical communication 
(Fig.~\ref{single} (c))
until they finish the computation.
Bob finally sends the final output of the quantum computation to Alice.
The whole protocol can be done in such a way that
Bob learns nothing about Alice's input, output, and algorithm.
Recently an efficient scheme for preparation of AKLT state with biphotons was proposed in~\cite{Darmawan} and
it seems that our single-server protocol can be adapted to such a scheme but we leave it as an open question what the best implementation is. 

\begin{figure}[htbp]
\begin{center}
\includegraphics[width=0.8\textwidth]{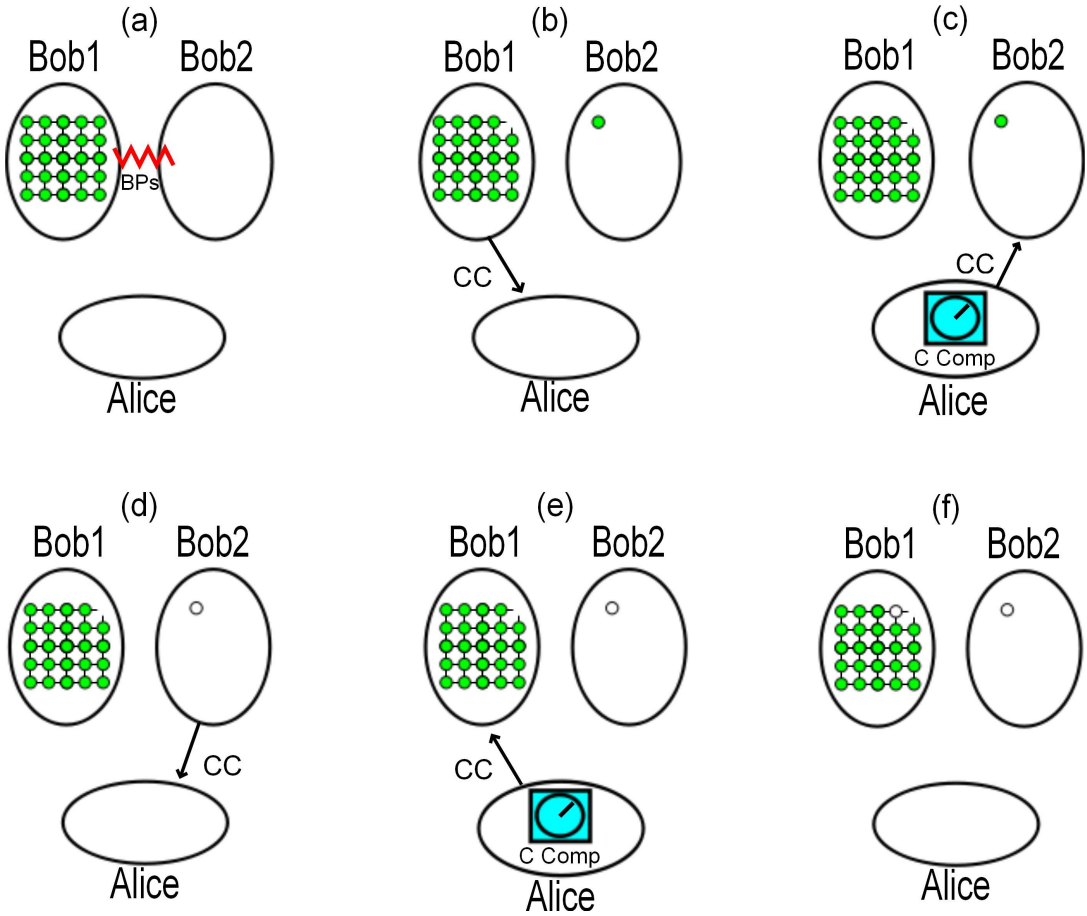}
\end{center}
\caption{(Color online.) 
The double-server protocol.
(a): 
Bob1 creates AKLT resource states.
Bob1 and Bob2 share Bell pairs (BPs).
(b): Bob1 adiabatically turns off the interaction between
some particles and others in 
his resource state,
and teleports these isolated particles to Bob2 by consuming Bell pairs.
Bob1 also sends Alice the result of the measurement in the teleportation.
(c): Alice calculates the angle in which particles should be measured
by using her classical computer (C Comp) and sends Bob2
the angle which is the sum of thus calculated angle and a random
angle.
(d): Bob2 performs the measurement in that angle
and sends the result of the measurement to Alice.
(e): Alice sends Bob1 the previous random angle.
(f): Bob1 implements the single-qubit rotation which
compensates the random rotation.
} 
\label{double_plain}
\end{figure}

As we argue later, the single-server protocol cannot
enjoy the energy-gap protection in condensed-matter systems,
since Bob cannot prepare any natural parent Hamiltonian without knowing hidden
angles in Dango states. Moreover, although Alice's quantum instrument is too primitive 
to perform the universal quantum
computation, and it does not seem to be hard to implement this
instrument in certain optical setups, it would be ideal if we can make Alice completely classical. Therefore, similar to \cite{blindcluster}, we consider an extension of the protocol to the double-server protocol.
In the double-server protocol (Fig.~\ref{double_plain}), there are three parties Alice, Bob1, and Bob2.
Alice, the client, has only a classical computer, whereas Bob1 and Bob2,
servers, have universal quantum computers.
Furthermore, Bob1 and Bob2 share many Bell pairs but they have no classical or quantum channel between them.
In the double-server protocol, Bob1 first creates AKLT resource 
states (Fig.~\ref{double_plain} (a)).
Bob1 next adiabatically turns off the interaction between
some particles and others in his resource state,
and teleports these particles to Bob2 by consuming Bell pairs. Bob1 sends Alice the result of the measurement in the teleportation 
through the classical channel (Fig.~\ref{double_plain} (b)). Note that due to the lack of any communication (classical or quantum) channels between Bob1 and Bob2, the teleportation procedure from Bobs' point of view can be seen as a usage of a totally mixed channel where only Alice knows how to correct the output of the channel. 
Next, Alice calculates
the angle in which particles should be measured by using
her classical computer, and
sends Bob2 the angle which is the sum of the calculated angle
plus a random angle
through the classical channel (Fig.~\ref{double_plain} (c)).
Bob2 performs the measurement in that angle
and sends the result of the measurement to Alice
(Fig.~\ref{double_plain} (d)).
Alice sends the previous random angle to Bob1 (Fig.~\ref{double_plain} (e)), 
and Bob1 does
the single-qubit rotation which compensates that random angle
(Fig.~\ref{double_plain} (f)).
Bob1 and Bob2 repeat this two-way classical communication with Alice until they finish the computation.
The whole protocol can be done in such a way that
two Bobs learn nothing about Alice's input, output, and algorithm. As we will argue later, the scheme of double servers proposed in \cite{blindcluster} where Bob1 prepares random states and Bob2 performs the measurement that compensates that random angle, will not lead to a ground state BQC protocol without affecting the privacy condition. While our new protocol can fully benefit from the energy-gap protection in condensed-matter systems, without revealing any information about Alice's secret computation.


This paper is organized as follows.
In Sec.~\ref{Sec:review}, 
we briefly review the concept of correlation space~\cite{Gross,Gross2,Gross3}
which is used throughout this paper 
in order to describe measurement-based quantum computation
on AKLT states.
We also briefly review 
the AKLT model and measurement-based 
quantum computing with AKLT resource.
In 
Sec.~\ref{Sec:single},
we explain the single-server protocol. 
We show the blindness of the single-server protocol 
in Sec.~\ref{Sec:single_proof}.
We also explain the double-server protocol
in Sec.~\ref{Sec:double}
and show its blindness
in Sec.~\ref{Sec:double_proof}. 
We finish with a discussion in Sec.~\ref{Sec:disc} which also highlights 
open problems.

\section{Quantum computation in correlation space}
\label{Sec:review}
In this section, we will briefly review the
concept of correlation space~\cite{Gross,Gross2,Gross3},
which is used throughout this paper.
For details see Refs.~\cite{Gross,Gross2,Gross3}.

Let us consider
the matrix-product state of a quantum state in
the $d^{N}$-dimensional Hilbert space
\begin{eqnarray*}
\sum_{l_1=1}^d...\sum_{l_N=1}^d\langle L|
A[l_N]...A[l_1]|R\rangle
|l_N\rangle\otimes...\otimes|l_1\rangle,
\end{eqnarray*}
where $|L\rangle$ and $|R\rangle$ are two-dimensional complex
vectors and $A$'s are two-dimensional complex matrices.
Let us assume that the first qudit of the matrix-product state
is projected onto 
\begin{eqnarray*}
|\theta,\phi\rangle=\cos\frac{\theta}{2}|0\rangle
+e^{i\phi}\sin\frac{\theta}{2}|1\rangle.
\end{eqnarray*}
Then, the matrix-product state becomes
\begin{eqnarray*}
\sum_{l_2=1}^d...\sum_{l_N=1}^d\langle L|
A[l_N]...A[l_2]A[\theta,\phi]|R\rangle
|l_N\rangle\otimes...\otimes|l_2\rangle\otimes|\theta,\phi\rangle,
\end{eqnarray*}
where
\begin{eqnarray*}
A[\theta,\phi]=\cos\frac{\theta}{2}A[0]+e^{-i\phi}\sin\frac{\theta}{2}A[1].
\end{eqnarray*}
If $A[0]$ and $A[1]$ are appropriately chosen in such a way that
$A[\theta,\phi]$ is unitary, we can ``simulate" the unitary
rotation $A[\theta,\phi]|R\rangle$ of $|R\rangle$
in the linear space where $|L\rangle$, $|R\rangle$, and $A$'s live.
This linear space is called ``correlation space"~\cite{Gross,Gross2,Gross3}.
In the general framework of measurement-based quantum computation,
which is called the computational tensor network~\cite{Gross,Gross2,Gross3},
universal quantum computation is performed in this correlation space.
Note that in the one-way quantum computing model, the correlation space
and the physical space are the same.
This separation between the correlation space and the physical space
will allow us to use many new resource states for measurement-based
quantum computing.
One such resource state is the AKLT state~\cite{MiyakeAKLT}.

Let us consider the one-dimensional open-boundary chain of $N$ 
qutrits.
The AKLT Hamiltonian~\cite{AKLT} is defined by
\begin{eqnarray*}
H_{AKLT}(\beta)\equiv\sum_{j=1}^{N-1}h_{j+1,j}(\beta),
\end{eqnarray*}
where
\begin{eqnarray*}
h_{j+1,j}(\beta)\equiv\frac{1}{2}[
{\mathbf S_{j+1}}\cdot{\mathbf S}_j
-\beta({\mathbf S_{j+1}}\cdot{\mathbf S}_j)^2]+\frac{1}{3}
\end{eqnarray*}
and
$\mathbf{S}_j\equiv(S_j^x,S_j^y,S_j^z)$ 
is the spin-1 operator on site $j$ defined by
\begin{eqnarray*}
S_j^x&\equiv&\frac{1}{\sqrt{2}}
\Bigg(
\begin{array}{ccc}
0&1&0\\
1&0&1\\
0&1&0
\end{array}
\Bigg),\\
S_j^y&\equiv&\frac{-i}{\sqrt{2}}
\Bigg(
\begin{array}{ccc}
0&1&0\\
-1&0&1\\
0&-1&0
\end{array}
\Bigg),\\
S_j^z&\equiv&
\Bigg(
\begin{array}{ccc}
1&0&0\\
0&0&0\\
0&0&-1
\end{array}
\Bigg).
\end{eqnarray*}
If $-1<\beta<1$, the system is in the gapped Haldane phase~\cite{Haldane}.
In particular, if $\beta=-1/3$,
Each $h_{j+1,j}(-1/3)$
is the projection operator onto the eigenspace of
the total spin operator of two spin-$1/2$ particles
corresponding to the eigenvalue 2.
For $\beta=-1/3$, the ground states are
called AKLT states~\cite{AKLT}
and
explicitly
written in the following matrix product form~\cite{AKLT,MiyakeAKLT}:
\begin{eqnarray}\label{eq:AKLT}
|AKLT^{N,L,R}\rangle\equiv
\frac{\sqrt{2}}{\sqrt{3^N}}
\sum_{l_1=1}^3...\sum_{l_N=1}^3\langle L|
A[l_N]...A[l_1]|R\rangle
|l_N\rangle\otimes...\otimes|l_1\rangle,
\end{eqnarray}
where 
\begin{eqnarray*}
|1\rangle&\equiv&-\frac{1}{\sqrt{2}}\Big(|S_z=+1\rangle-|S_z=-1\rangle\Big),\\
|2\rangle&\equiv&\frac{1}{\sqrt{2}}\Big(|S_z=+1\rangle+|S_z=-1\rangle\Big),\\
|3\rangle&\equiv&|S_z=0\rangle,
\end{eqnarray*}
$|S_z=k\rangle$ ($k\in\{-1,0,+1\}$) are eigenvectors of the $z$-component
$S_z$ of the spin-1 operator,
$|L\rangle$ and $|R\rangle$ are two-dimensional complex vectors,
and $\{A[1],A[2],A[3]\}$ are $2\times2$ matrices
defined by
\begin{eqnarray*}
A[1]&\equiv&X,\\
A[2]&\equiv&XZ,\\
A[3]&\equiv&Z.
\end{eqnarray*}
Here, 
$X$ and
$Z$
are Pauli operators over qubits. 
Note that the ground states of the AKLT Hamiltonian
are four-fold degenerate, and each ground state
is specified with $|L\rangle$ and $|R\rangle$,
which represent two qubit edge states.
The AKLT states are frustration free, since  
\begin{eqnarray*}
h_{j+1,j}(-1/3)|AKLT^{N,L,R}\rangle=0
\end{eqnarray*}
for any $|L\rangle$ and $|R\rangle$, and for all $j=1,...,N-1$.

The AKLT model has long been studied 
in condensed matter physics since it can
describe the one-dimensional Haldane phase~\cite{Haldane} of
a qutrit chain, which exhibits the spectral gap~\cite{AKLT}, the diluted
antiferromagnetic order detected by the string order parameter~\cite{string},
and the effective spin-1/2 degree of freedom, namely the edge state,
appearing on the boundary of the chain~\cite{edge}.
Furthermore, the AKLT model has recently been attracting much attentions
in the field of quantum information, because of its connections
to the matrix product 
representation~\cite{Garcia,Verstraete_review,Cirac_review},
the localizable entanglement~\cite{localizable,Delgado},
and the measurement-based quantum 
computation~\cite{MiyakeAKLT,Miyakeholographic,Gross,Miyake2dAKLT,Wei}.
Indeed, it was shown in Refs.~\cite{MiyakeAKLT,Miyakeholographic,Miyake2dAKLT,Wei} 
that
the measurement-based quantum computation is possible
on the AKLT chains or other ground states
in the gapped Haldane phase ($-1<\beta<1$).

We briefly review universal measurement-based quantum computation
on AKLT states~\cite{MiyakeAKLT}. 
Assume that
a qutrit of Eq.~(\ref{eq:AKLT}) is measured in the basis
${\mathcal M(\phi)}
=\{|\alpha(\phi)\rangle,|\beta(\phi)\rangle,|\gamma\rangle\}$,
where
\begin{eqnarray*}
|\alpha(\phi)\rangle&=&\frac{1+e^{i\phi}}{2}|1\rangle
+\frac{1-e^{i\phi}}{2}|2\rangle,\\
|\beta(\phi)\rangle&=&\frac{1-e^{i\phi}}{2}|1\rangle
+\frac{1+e^{i\phi}}{2}|2\rangle,\\
|\gamma\rangle&=&|3\rangle.
\end{eqnarray*}
Then, following operations are implemented in the correlation space
according to the measurement result~\cite{MiyakeAKLT}.
\begin{eqnarray*}
|\alpha(\phi)\rangle&:&~~~Xe^{i\phi Z/2},\\
|\beta(\phi)\rangle&:&~~~XZe^{i\phi Z/2},\\
|\gamma\rangle&:&~~~Z.
\end{eqnarray*}

Now assume that the unitary operation
\begin{eqnarray*}
V=|3\rangle\langle1|+|1\rangle\langle2|+|2\rangle\langle3|
\end{eqnarray*}
is applied on a qutrit and that qutrit is measured in the basis
${\mathcal M(\phi)}$.
Then,
following operations are implemented in the correlation space
according to the measurement result~\cite{MiyakeAKLT}.
\begin{eqnarray*}
|\alpha(\phi)\rangle&:&~~~XZe^{-i\phi X/2},\\
|\beta(\phi)\rangle&:&~~~Ze^{-i\phi X/2},\\
|\gamma\rangle&:&~~~X.
\end{eqnarray*}

In this way, any single-qubit $Z$ and $X$ rotations are possible up
to some Pauli byproducts.
By appropriately changing the sign of $\phi$ 
(adapting measurements of the physical qutrits), these byproducts
can be moved forward so that they are corrected in the final
stage of the computation over the correlation space.

\section{Single-sever Blind Quantum Computing Protocol}
\label{Sec:single}

As said before, in the single-server BQC protocol (Fig.~\ref{single}), Alice has a classical computer and a quantum instrument that emits random four-qubit states called
``Dango states". Depending on the desired computation and the input size, Alice will send ($2\times N\times M$) Dango states directly to Bob through a one-way quantum channel that they initially share. Bob stores all of them in his quantum memory to creates the resource state, called ``rotated AKLT states''. The procedure of preparing such an initial state is explained in the Blind state preparation Subsection below.

Next, Alice calculates the angle in which a particle of a rotated AKLT state should be measured. Recall that this is a qutrit measurement that will induce a qubit operation over the correlation space. Moreover the calculated angle should compensate for the initial random rotation of the Dango states and byproduct operation of the previous measurement. Finally an additional random rotation will be added to hide the true result of the measurement from Bob. Bob performs the measurement according to Alice's information (sent via a classical channel to him), and returns the result of the measurement to Alice. They repeat this two-way classical communication until they finish the computation. Bob finally sends the final output of the quantum computation to Alice. The exact protocol is given in the Blind computation Subsection below where we describe how a blind arbitrary $X$ and $Z$ rotation in the correlation space can be performed. Next we describe how two-qubit entangling operation of $CZ$ can be performed in regular places. The rotation operators are also performed in regular interval, and hence the overall structure of the actual underlying computation remains hidden to Bob. These set of operators define a universal set of gates for quantum computing.

\subsection{Blind state preparation}
\begin{figure}[htbp]
\begin{center}
\includegraphics[width=0.6\textwidth]{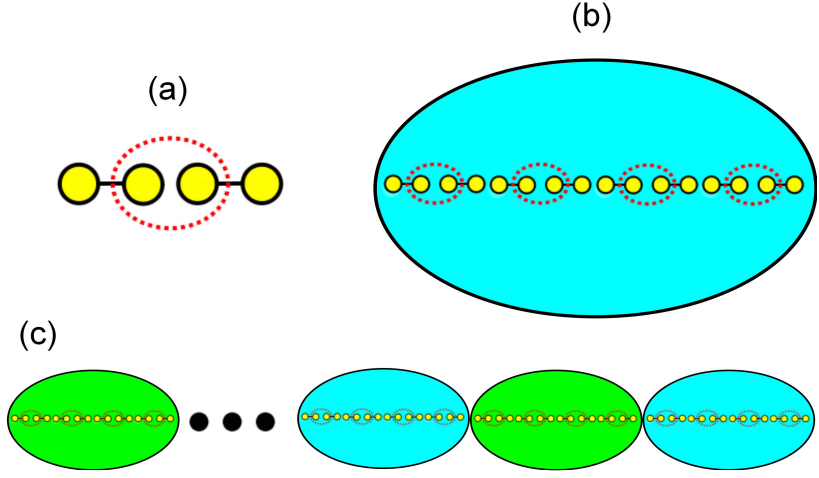}
\end{center}
\caption{(Color online.) 
(a): A Dango state.
Two yellow circles connected by a bond are qubits which consist
of the Bell state. The operator $(I-|\eta_1\rangle\langle\eta_1|)T_{Z/X}(\xi)$ 
acts on two qubits in
the red dotted circle.
(b): A $Z$-Dango chain for $n=4$.
(c): A Combo chain $|C_b\rangle$. $Z$-Dango chains are colored in blue,
whereas $X$-Dango chains are colored in green.
} 
\label{blob}
\end{figure}

Denote the Bell basis by
\begin{eqnarray*}
|\eta_1\rangle&\equiv&
\frac{1}{\sqrt{2}}
\Big(|0\rangle\otimes|0\rangle+|1\rangle\otimes|1\rangle\Big),\\
|\eta_2\rangle&\equiv&
\frac{1}{\sqrt{2}}
\Big(|0\rangle\otimes|0\rangle-|1\rangle\otimes|1\rangle\Big),\\
|\eta_3\rangle&\equiv&
\frac{1}{\sqrt{2}}
\Big(|1\rangle\otimes|0\rangle+|0\rangle\otimes|1\rangle\Big),\\
|\eta_4\rangle&\equiv&
\frac{1}{\sqrt{2}}
\Big(|1\rangle\otimes|0\rangle-|0\rangle\otimes|1\rangle\Big).
\end{eqnarray*}

The full state preparation is described in Protocol 1. 
Assume Alice's desired computation is composed of a sequence of  $X$ and $Z$-rotations and $CZ$ operations. Depending on the number of the required operators and the size of the input, Alice will 
choose integer values $N$ and $M$.
Then Alice's quantum instrument emits $N\times M$ ``$Z$-Dango states"  and 
$N\times M$ ``$X$-Dango states" defined as (see Fig.~\ref{blob} (a))
\begin{eqnarray*}
|D_Z(\xi_{a,b}^Z)\rangle\equiv
(I\otimes (I-|\eta_1\rangle\langle\eta_1|)\otimes I)
(I\otimes T_Z(\xi_{a,b}^Z)\otimes I)
|\eta_1\rangle\otimes|\eta_1\rangle, \\
|D_X(\xi_{a,b}^X)\rangle\equiv
(I\otimes (I-|\eta_1\rangle\langle\eta_1|)\otimes I)
(I\otimes T_X(\xi_{a,b}^X)\otimes I)
|\eta_1\rangle\otimes|\eta_1\rangle,
\end{eqnarray*}
where 
$(a,b)\in\{1,...,N\}\times\{1,...,M\}$, 
$\xi_{a,b}^{Z/X}\in{\mathcal A} \equiv \Big\{\frac{k\pi}{4}\Big|k=0,...,7\Big\}$ are independently and uniformly distributed
random numbers which are secret to
Bob, 
and the two qubit operators $T_Z(\xi_{a,b}^Z)$ 
and $T_X(\xi_{a,b}^X)$ are defined by
\begin{eqnarray*}
T_Z(\xi_{a,b}^Z)&\equiv&
|00\rangle\langle00|
+e^{i\xi_{a,b}^Z}|01\rangle\langle01|
+|10\rangle\langle10|
+|11\rangle\langle11|,\\
T_X(\xi_{a,b}^X)&\equiv&
\Big(\frac{1+e^{i\xi_{a,b}^X}}{2}|\eta_2\rangle
+\frac{1-e^{i\xi_{a,b}^X}}{2}|\eta_4\rangle\Big)\langle \eta_2|
+\Big(\frac{1-e^{i\xi_{a,b}^X}}{2}|\eta_2\rangle
+\frac{1+e^{i\xi_{a,b}^X}}{2}|\eta_4\rangle\Big)\langle \eta_4|
+|\eta_1\rangle\langle \eta_1|
+|\eta_3\rangle\langle \eta_3|.
\end{eqnarray*}
Alice sends all these Dango states to Bob,
and records all $\{\xi_{a,b}^{Z/X}\}$ for the later 
use~\cite{Bobcandofiltering}. 
Bob arranges all the Dango states in a lattice with $2N$ columns and
$M$ rows.

\begin{figure}[htbp]
\begin{center}
\includegraphics[width=0.8\textwidth]{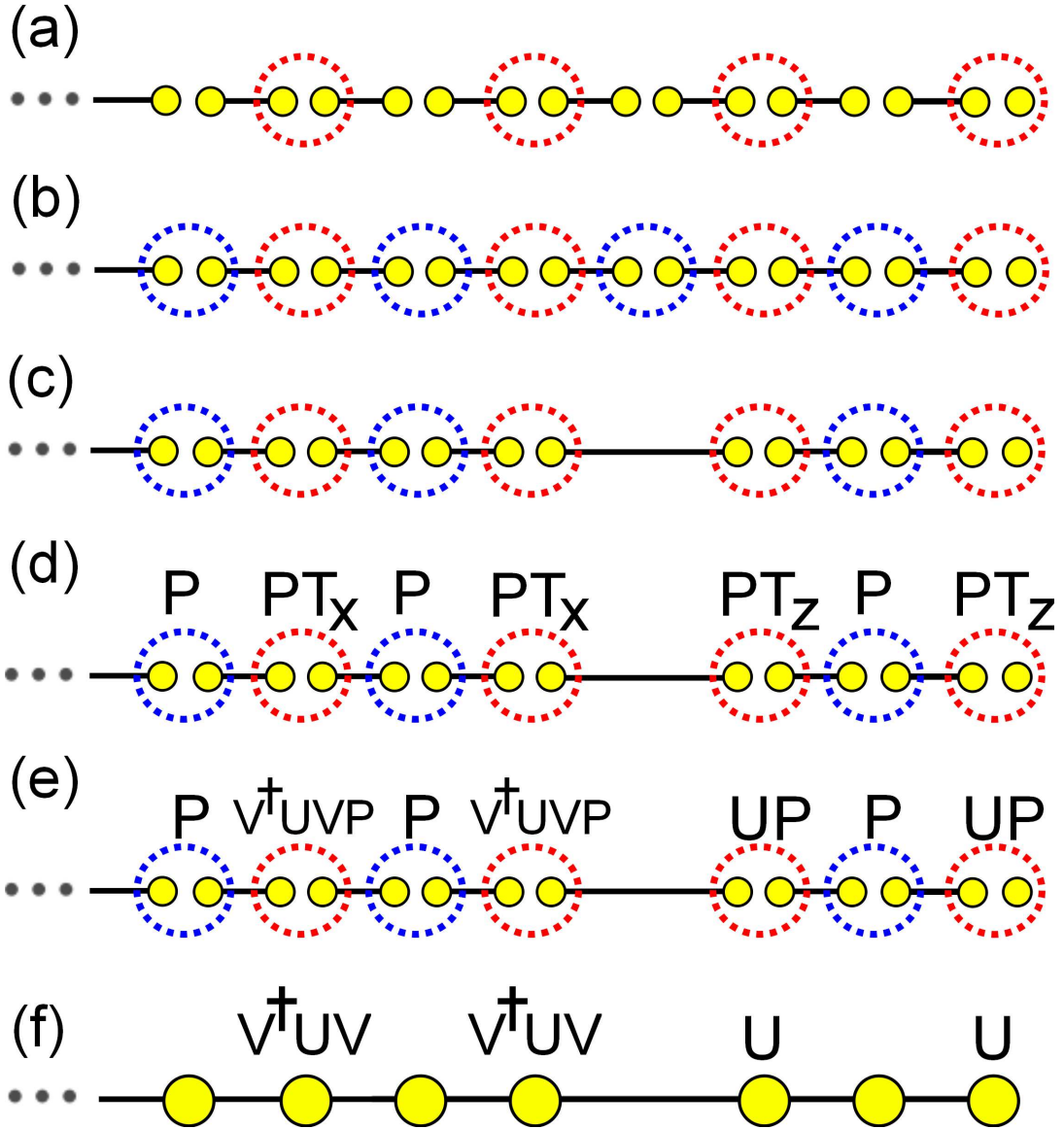}
\end{center}
\caption{(Color online.) 
(a): A chain $|C\rangle$. 
(b): The filtering operation is applied
to every pair of qubits which is indicated
by a blue dotted circle. 
(c): $|\eta_1\rangle\langle\eta_1|$ is realized on the
seventh and eighth qubits.
(d): The PEPS $P$ is applied to every pair of two qubits.
(e): It is equivalent to (d).
(f): It is equivalent to (e).
Now large yellow circles are qutrits.
} 
\label{spin1creation}
\end{figure}

Alice chooses a parameter $n< N$. We call a collection of $n$ Dango states, 
sent by Alice to be kept in Bob's memory, `` $(k,b)$th $Z$-Dango chain states" 
or ``$(k,b)$th $X$-Dango chain states" defines as (see Fig.~\ref{blob} (b))
\begin{eqnarray*}
|B^Z_{k,b}\rangle & \equiv &
\bigotimes_{j=1}^n|D_Z(\xi_{(k-1)n+j,b}^Z)\rangle,\\ 
|B^X_{k,b}\rangle &\equiv &
\bigotimes_{j=1}^n|D_X(\xi_{(k-1)n+j,b}^X)\rangle,
\end{eqnarray*}
where $k=1,...,N/n$ and $b=1,...,M$.
A $Z$-Dango chain state is used for the implementation of a single-qubit $Z$-rotation whereas
an $X$-Dango chain state is used for the implementation a single-qubit $X$-rotation. However, to hide the actual structure of the computation Alice will work with a regular one-dimensional chain, called ``Combo chain state'' $|C_b\rangle$, 
composed of $N/n$ $Z$-Dango chain states and $N/n$ $X$-Dango chain states 
(Fig.~\ref{blob} (c)) 
with two-edge qubits projected on
$|R^*\rangle$ and $|L\rangle$ states, respectively 
(Fig. \ref{spin1creation} (a)): 
\begin{eqnarray*}
|C_b\rangle\equiv
\langle R^*|\langle L| 
\Big(|B^X_{N/n,b}\rangle\otimes|B^Z_{N/n,b}\rangle
\otimes...\otimes
|B^X_{2,b}\rangle\otimes|B^Z_{2,b}\rangle\otimes
|B^X_{1,b}\rangle\otimes|B^Z_{1,b}\rangle\Big)
\end{eqnarray*}
($b=1,...,M$).
Here, $|L\rangle=(|0\rangle+i|1\rangle)/\sqrt{2}$ 
and $|R\rangle=|0\rangle$.
However, $|R\rangle$ could be any arbitrary state
depending on Alice's desired input using the uploading 
method~\cite{uploading}.

Now in order to entangle qubits of
the Combo chain to create the desired resource state,
Bob will perform the following operators. Let us define the PEPS operation to be~\cite{Garcia,Verstraete_review,Cirac_review}
\begin{eqnarray*}
P\equiv\frac{1}{\sqrt{2}}
\sum_{l=1}^3\sum_{i=0}^1\sum_{j=0}^1A_{i,j}[l]|l\rangle
\langle i|\otimes\langle j|,
\end{eqnarray*}
which creates a qutrit from two qubits,
where $A_{i,j}[l]$ is $(i,j)$-element of the matrix $A[l]$. Consider the following unitary operators acting on a qutrit
\begin{eqnarray*}
U(\xi_{a,b}^{Z/X})&\equiv&
\Big(\frac{1+e^{i\xi_{a,b}^{Z/X}}}{2}|1\rangle
+\frac{1-e^{i\xi_{a,b}^{Z/X}}}{2}|2\rangle\Big)\langle1|
+\Big(\frac{1-e^{i\xi_{a,b}^{Z/X}}}{2}|1\rangle
+\frac{1+e^{i\xi_{a,b}^{Z/X}}}{2}|2\rangle\Big)\langle2|
+|3\rangle\langle3|, \\
V&\equiv&|3\rangle\langle1|
+|1\rangle\langle2|
+|2\rangle\langle3|.
\end{eqnarray*}
It is easy to verify that 
\begin{eqnarray*}
PT_Z(\xi_{a,b}^{Z/X})&=&U(\xi_{a,b}^{Z/X})P,\\
PT_X(\xi_{a,b}^{Z/X})&=&V^\dagger U(\xi_{a,b}^{Z/X})VP.
\end{eqnarray*}

Bob has to apply the filtering operation
$I-|\eta_1\rangle\langle\eta_1|$.
In order to do so, he performs the measurement
$\{|\eta_1\rangle\langle\eta_1|,I-|\eta_1\rangle\langle\eta_1|\}$
to every pair of two qubits in the Combo chain which is
specified by a dotted blue circle in Fig.~\ref{spin1creation} (b). 
If $|\eta_1\rangle\langle\eta_1|$
is realised, two qubits are just removed from the chain
(Fig.~\ref{spin1creation} (c)).  
Next Bob applies the PEPS operation $P$ to each pair of two qubits
in order to obtain qutrits (Fig.~\ref{spin1creation} (d), (e), (f))~\cite{inBob'sbrain}.
This PEPS operation is done deterministically since 
$I-|\eta_1\rangle\langle\eta_1|$
is already applied to every pair of qubits. Therefore 
Bob has created a new one-dimensional chain of qutrits 
(Fig.~\ref{spin1creation} (f)) called ``rotated AKLT state'': 
\begin{eqnarray*}
|RAKLT_{b}^{2N,L,R}(\{\xi_{a,b}^{Z/X}\})\rangle=
\mathcal{U}_b(\{\xi_{a,b}^{Z/X}\})|AKLT^{2N,L,R}\rangle,
\end{eqnarray*}
where
\begin{eqnarray*}
{\mathcal U}_b(\{\xi_{a,b}^{Z/X}\})&\equiv&
\Big\{
V^\dagger U(\xi_{N,b}^X)V
\otimes...
\otimes V^\dagger U(\xi_{N-n+1,b}^X)V\Big\}
\otimes
\Big\{U(\xi_{N,b}^Z)\otimes...
\otimes U(\xi_{N-n+1,b}^Z)\Big\}\\
&&\vdots\\
&&\otimes
\Big\{V^\dagger U(\xi_{n+n,b}^X)V\otimes...
\otimes V^\dagger U(\xi_{n+1,b}^X)V\Big\}\otimes
\Big\{U(\xi_{n+n,b}^Z)\otimes...
\otimes U(\xi_{n+1,b}^Z)\Big\}\\
&&\otimes
\Big\{V^\dagger U(\xi_{n,b}^X)V\otimes...
\otimes V^\dagger U(\xi_{1,b}^X)V\Big\}
\otimes
\Big\{U(\xi_{n,b}^Z)\otimes...\otimes U(\xi_{1,b}^Z)\Big\}
\end{eqnarray*} 
(for simplicity, we have assumed that all filterings give $|\eta_1\rangle\langle\eta_1|$).
We call a qutrit which is rotated by 
$U(\xi_{a,b}^Z)$ ``$Z$-prerotated qutrit" and a
qutrit which is rotated by $V^\dagger U(\xi_{a,b}^X)V$ 
``$X$-prerotated qutrit". Other qutrits are called ``plain qutrits". 
The ``$(k,b)$th $Z/X$-prerotated AKLT subsystem" is defined 
to be the set of $Z/X$-prerotated qutrits in $b$th
prerotated AKLT chain
corresponding to particles of $(k,b)$th $Z/X$-Dango chain.

\begin{algorithm}
\caption{\bf 1: Blind State Preparation}
\label{prot:state}
\begin{itemize} 
\item Alice sends Bob parameter values $N$, $M$ and $n<N$.
\item Alice sends Bob $N\times M$ many $Z$-Dango states $|D_Z(\xi^Z_{a,b})\rangle$ where $(a,b)\in\{1,...,N\}\times\{1,...,M\}$.
\item Alice sends Bob $N\times M$ many $X$-Dango states $|D_X(\xi^X_{a,b})\rangle$ where $(a,b)\in\{1,...,N\}\times\{1,...,M\}$.
\item Bob arranges the received $Z$-Dango states in $M$ rows of $N/n$ $Z$-Dango chains $|B^Z_{k,b}\rangle$ where $k=1,2,...,N/n$, $b=1,...,M$.  
\item Bob arranges the received $X$-Dango states in $M$ rows of $N/n$ $X$-Dango chains $|B^X_{k,b}\rangle$ where $k=1,2,...,N/n$, $b=1,...,M$.  
\item Bob arranges the Dango chains in $M$ rows of Combo chains $|C_{b}\rangle \equiv |B^X_{N/n,b}\rangle\otimes|B^Z_{N/n,b}\rangle \otimes...\otimes|B^X_{1,b}\rangle\otimes|B^Z_{1,b}\rangle$.
\item Bob applies filtering and PEPS operators to create $M$ rows of rotated AKLT states $|RAKLT_{b}^{2N,L,R}(\{\xi_{a,b}^{Z/X}\})\rangle$, where $b=1,...,M$.
\end{itemize}
See also Fig.~\ref{spin1creation}.
\end{algorithm}

\subsection{Blind computation protocol}

A single blind $Z$-rotation is performed using a $Z$-Dango chain state. Let us assume that Alice wants to perform
the $Z$-rotation $\exp\Big[\frac{iZ}{2}\theta_{k,b}^Z\Big]$ with $\theta^Z_{k,b}\in {\mathcal A}$, using
the $(k,b)$th $Z$-Dango chain~\cite{pi8_universal}. The step by step operation is given in Protocol 2. 
Note that we implement the desired $Z$-rotation
using qutrit measurements. However, the third outcome $|\gamma\rangle$
leads to the failure as it implements only the trivial Pauli $Z$. 
The probability that Alice fails to implement her desired $Z$-rotation in a single $Z$-Dango chain is $1/3^n$, which is small for sufficiently large $n$. Similarly a blind $X$ rotation could be applied 
(see Protocol 3) where again, the probability that Alice fails to implement her desired $X$-rotation in a single 
$X$-Dango chain is $1/3^n$.
Note that these protocols are designed in such a way to
follow the construction of the AKLT computation of Sec.~\ref{Sec:review},
while canceling the prerotation that was added to the resource state
for the purpose of blindness.

\begin{algorithm}
\caption{\bf 2: Blind $Z$ Rotation}
\label{prot:Zrot}

Initially the flag parameter (known to both Alice and Bob) is set $\tau=1$.  For $j=1 \cdots n$ Alice and Bob perform the following steps.
\begin{itemize}
\item[\bf (I)]
Alice sends Bob the angle 
\begin{eqnarray*}
\phi^Z_{(k-1)n+j,b}\equiv\tau\theta^Z_{k,b}+\xi^Z_{(k-1)n+j,b}
+\pi r^Z_{(k-1)n+j,b}~~~(\mbox{mod}~2\pi), 
\end{eqnarray*}
where 
$r^Z_{(k-1)n+j,b}\in\{0,1\}$ is a random number which is secret to Bob.
If there is the $X$ byproduct before this step, 
$\theta^Z_{k,b}$ should be replaced with
$-\theta^Z_{k,b}$ in order to compensate this byproduct operator.
However, $Z$ byproduct commutes trivially with the operation implemented
in the correlation space, and therefore
it can be corrected at the end of computation.
\item[\bf (II)]
Bob measures  
the $j$th $Z$-prerotated qutrit  
of the $(k,b)$th $Z$-prerotated AKLT subsystem
in the basis
\begin{eqnarray*}
\mathcal{M}(\phi^Z_{(k-1)n+j,b})\equiv
\{|\alpha(\phi^Z_{(k-1)n+j,b})\rangle,
|\beta(\phi^Z_{(k-1)n+j,b})\rangle,
|\gamma\rangle\},
\end{eqnarray*}
where
\begin{eqnarray*}
|\alpha(\phi^Z_{(k-1)n+j,b})\rangle
&\equiv&
\frac{1+\exp\Big[i\phi^Z_{(k-1)n+j,b}\Big]}{2}|1\rangle
+\frac{1-\exp\Big[i\phi^Z_{(k-1)n+j,b}\Big]}{2}|2\rangle,\\
|\beta(\phi^Z_{(k-1)n+j,b})\rangle
&\equiv&
\frac{1-\exp\Big[i\phi^Z_{(k-1)n+j,b}\Big]}{2}|1\rangle
+\frac{1+\exp\Big[i\phi^Z_{(k-1)n+j,b}\Big]}{2}|2\rangle,\\
|\gamma\rangle&\equiv&|3\rangle.
\end{eqnarray*}
and sends the result to Alice. 
\begin{itemize}
\item[$\bullet$]
If the measurement result is
$|\alpha(\phi^Z_{(k-1)n+j,b})\rangle$,
\begin{eqnarray*}
R^\alpha_Z(\tau\theta^Z_{k,b},r^Z_{(k-1)n+j,b})
\equiv
\exp\Big[\frac{-i\tau\theta_{k,b}^Z}{2}\Big]
XZ^{r^Z_{(k-1)n+j,b}}\exp\Big[\frac{iZ}{2}\tau\theta^Z_{k,b}\Big]
\end{eqnarray*}
is implemented in the correlation space and Alice sets $\tau=0$.
\item[$\bullet$]
If the measurement result is
$|\beta(\phi^Z_{(k-1)n+j,b})\rangle$,
\begin{eqnarray*}
R^\beta_Z(\tau\theta^Z_{k,b},r^Z_{(k-1)n+j,b})
\equiv
\exp\Big[\frac{-i\tau\theta_{k,b}^Z}{2}\Big]XZ^{r^Z_{(k-1)n+j,b}+1}
\exp\Big[\frac{iZ}{2}\tau\theta^Z_{k,b}\Big]
\end{eqnarray*}
is implemented in the correlation space and Alice sets $\tau=0$.
\item[$\bullet$]
If the measurement result is
$|\gamma\rangle$, 
$Z$
is implemented in the correlation space.
\end{itemize}
The probability of obtaining 
each result
is $1/3$.
\item[\bf (III)]
Bob does the measurement $\{|1\rangle,|2\rangle,|3\rangle\}$
on the next plain qutrit if any.
\end{itemize}
\end{algorithm}

\begin{algorithm}
\caption{\bf 3: Blind $X$ Rotation}
\label{prot:Xrot}

Initially the flag parameter (known to both Alice and Bob) is set $\tau=1$.  For $j=1 \cdots n$ Alice and Bob perform the following steps.

\begin{itemize}

\item[\bf (I)] Alice sends Bob the angle 
\begin{eqnarray*}
\phi^X_{(k-1)n+j,b}\equiv
\tau\theta^X_{k,b}+\xi^X_{(k-1)n+j,b}+r^X_{(k-1)n+j,b}\pi~~~(\mbox{mod}~2\pi), 
\end{eqnarray*}
where 
$r^X_{(k-1)n+j,b}\in\{0,1\}$ is a random number which is secret to Bob.
If there is the $Z$ byproduct operator before this step, 
$\theta^X_{k,b}$ should be replaced with
$-\theta^X_{k,b}$ in order to compensate this byproduct operator.
However, $X$ byproduct commutes trivially with the operation implemented
in the correlation space, and therefore
it can be corrected at the end of computation.

\item[\bf(II)] Bob applies $V$ on the $j$th $X$-prerotated qutrit of the 
$(k,b)$th
$X$-prerotated AKLT subsystem,
and does the measurement in the basis 
\begin{eqnarray*}
\mathcal{M}(\phi^X_{(k-1)n+j,b})\equiv
\{|\alpha(\phi^X_{(k-1)n+j,b})\rangle,
|\beta(\phi^X_{(k-1)n+j,b})\rangle,
|\gamma\rangle\},
\end{eqnarray*}
where
\begin{eqnarray*}
|\alpha(\phi^X_{(k-1)n+j,b})\rangle
&\equiv&
\frac{1+\exp\Big[i\phi^X_{(k-1)n+j,b}\Big]}{2}|1\rangle
+\frac{1-\exp\Big[i\phi^X_{(k-1)n+j,b}\Big]}{2}|2\rangle,\\
|\beta(\phi^X_{(k-1)n+j,b})\rangle
&\equiv&
\frac{1-\exp\Big[i\phi^X_{(k-1)n+j,b}\Big]}{2}|1\rangle
+\frac{1+\exp\Big[i\phi^X_{(k-1)n+j,b}\Big]}{2}|2\rangle,\\
|\gamma\rangle&\equiv&|3\rangle.
\end{eqnarray*}
and sends the result to Alice. 
\begin{itemize}
\item[$\bullet$]
If the measurement result is
$|\alpha(\phi^X_{(k-1)n+j,b})\rangle$,
\begin{eqnarray*}
R_X^\alpha(\tau\theta^X_{k,b},r^X_{(k-1)n+j,b})\equiv
\exp\Big[\frac{-i\tau\theta_{k,b}^X}{2}\Big]
X^{r^X_{(k-1)n+j,b}+1}Z\exp\Big[\frac{-iX}{2}\tau\theta^X_{k,b}\Big]
\end{eqnarray*}
is implemented in the correlation space and Alice sets $\tau=0$.
\item[$\bullet$]
If the measurement result is
$|\beta(\phi^X_{(k-1)n+j,b})\rangle$,
\begin{eqnarray*}
R_X^\beta(\tau\theta^X_{k,b},r^X_{(k-1)n+j,b})\equiv
\exp\Big[\frac{-i\tau\theta_{k,b}^X}{2}\Big]X^{r^X_{(k-1)n+j,b}}Z
\exp\Big[\frac{-iX}{2}\tau\theta^X_{k,b}\Big]
\end{eqnarray*}
is implemented in the correlation space and Alice sets $\tau=0$.
\item[$\bullet$]
If the measurement result is
$|\gamma\rangle$,
$X$
is implemented.
\end{itemize}
The probability of obtaining each result is $1/3$.
\item[\bf (III)] Bob performs the measurement $\{|1\rangle,|2\rangle,|3\rangle\}$
on the next plain qutrit if any.
\end{itemize}
\end{algorithm}

Let us finally explain how the two-qubit operation of controlled-$Z$ ($CZ$) is performed. In order to perform $CZ$ gates blindly,
$CZ$ gates are periodically implemented 
with the period that is
independent 
of Alice's input and the algorithm. In this case,
Bob learns nothing from the period.
Because of the periodic implementation
of $CZ$ gates, Alice sometimes experiences
an unwanted $CZ$ gate.
However,
Alice can cancel the effect of an
unwanted $CZ$ gate by implementing
the trivial identity operation (plus Pauli byproduct operations) 
until she arrives at the next $CZ$ gate which
cancels the previous one. This is possible due to the following commutation rules.
\begin{eqnarray*}
CZ(I\otimes X)CZ&=&Z\otimes X,\\
CZ(I\otimes Z)CZ&=&I\otimes Z,\\
CZ(X\otimes I)CZ&=&X\otimes Z,\\
CZ(Z\otimes I)CZ&=&Z\otimes I.
\end{eqnarray*}
In Protocol 4,
we explain how to implement
the $CZ$ gate plus $Z$-rotation
\begin{eqnarray*}
\Big(\exp\Big[\frac{iZ}{2}\theta^Z_{k,b}\Big]
\otimes\exp\Big[\frac{iZ}{2}\theta^Z_{k',b'}\Big]\Big)
CZ
\end{eqnarray*}
between $(k,b)$th and $(k',b')$th $Z$-prerotated AKLT subsystems,
where $\theta^Z_{k,b},\theta^Z_{k',b'}\in{\mathcal A}$.
Note that the local $Z$ rotations are required to cancel
the prerotations of qutrits.
The probability that Alice fails to implement
the CZ gate in this algorithm is $(5/9)^n$, which is small
for large $n$.

\begin{algorithm}
\caption{\bf 4: Controlled-Z followed by Blind $Z$-rotations}
\label{prot:CZ}
Initially the flag parameters (known to both Alice and Bob) are set $\tau=1$, $\tau'=1$ and $\epsilon=1$.  For $j=1 \cdots n$  Alice and Bob perform the following steps.
\begin{itemize}

\item[\bf (I)]
If $\epsilon = 0,$ skip this step.
Bob applies
the unitary operation 
\begin{eqnarray*}
W&\equiv&
\frac{|1,1\rangle+|1,2\rangle+|2,1\rangle-|2,2\rangle}{2}\langle1,1|\\
&&+\frac{|1,1\rangle+|1,2\rangle-|2,1\rangle+|2,2\rangle}{2}\langle1,2|\\
&&+\frac{|1,1\rangle-|1,2\rangle+|2,1\rangle+|2,2\rangle}{2}\langle2,1|\\
&&+\frac{-|1,1\rangle+|1,2\rangle+|2,1\rangle+|2,2\rangle}{2}\langle2,2|\\
&&+|1,3\rangle\langle 1,3|
+|2,3\rangle\langle 2,3|
+|3,1\rangle\langle 3,1|
+|3,2\rangle\langle 3,2|
+|3,3\rangle\langle 3,3|
\end{eqnarray*}
between $j$th $Z$-prerotated qutrit of $(k,b)$th $Z$-prerotated AKLT subsystem and $j$th $Z$-prerotated qutrit of $(k',b')$th
$Z$-prerotated AKLT subsystem. 
\item[\bf (II)]
Alice sends Bob the angles
\begin{eqnarray*}
\phi^Z_{(k-1)n+j,b}&=&\tau\theta^Z_{k,b}+\xi^Z_{(k-1)n+j,b}+r^Z_{(k-1)n+j,b}
\pi~~(\mbox{mod}~2\pi),\\
\phi^Z_{(k'-1)n+j,b'}&=&\tau'\theta^Z_{k',b'}
+\xi^Z_{(k'-1)n+j,b'}+r^Z_{(k'-1)n+j,b'}\pi~~(\mbox{mod}~2\pi),
\end{eqnarray*}
where 
$r^Z_{(k-1)n+j,b},
r^Z_{(k'-1)n+j,b'}\in\{0,1\}$ are random numbers.
If there is any byproduct which contains $X$ before this step, the
sign of $\theta^Z_{k,b}$
or $\theta^Z_{k',b'}$ should be appropriately changed.
However, $Z$ byproduct commutes trivially with the operation implemented
in the correlation space, and therefore
it can be corrected at the end of computation.

\item[\bf (III)]
Bob does
the measurement ${\mathcal M}(\phi^Z_{(k-1)n+j,b})$ (the same
as that of Protocol 1) on
the $j$th $Z$-prerotated qutrit of the $(k,b)$th $Z$-prerotated AKLT subsystem and the measurement
${\mathcal M}(\phi^Z_{(k'-1)n+j,b'})$ on
the $j$th $Z$-prerotated qutrit of the $(k',b')$th $Z$-prerotated AKLT subsystem. 
The operation implemented in the correlation space is summarized
as follows:
\begin{itemize}

\item[$\bullet$]
$|\alpha(\phi^Z_{(k-1)n+j,b})\rangle\otimes
|\alpha(\phi^Z_{(k'-1)n+j,b'})\rangle$:
$\Big[R_Z^\alpha(\tau\theta^Z_{k,b},r^Z_{(k-1)n+j,b})
\otimes R_Z^\alpha(\tau'\theta^Z_{k',b'},r^Z_{(k'-1)n+j,b'})
\Big]CZ^\epsilon$

\item[$\bullet$]
$|\alpha(\phi^Z_{(k-1)n+j,b})\rangle\otimes
|\beta(\phi^Z_{(k'-1)n+j,b'})\rangle$:
$\Big[R_Z^\alpha(\tau\theta^Z_{k,b},r^Z_{(k-1)n+j,b})
\otimes R_Z^\beta(\tau'\theta^Z_{k',b'},r^Z_{(k'-1)n+j,b'})
\Big]CZ^\epsilon$

\item[$\bullet$]
$|\alpha(\phi^Z_{(k-1)n+j,b})\rangle\otimes
|\gamma\rangle$:
$R_Z^\alpha(\tau\theta^Z_{k,b},r^Z_{(k-1)n+j,b})
\otimes Z
$

\item[$\bullet$]
$|\beta(\phi^Z_{(k-1)n+j,b})\rangle\otimes
|\alpha(\phi^Z_{(k'-1)n+j,b'})\rangle$:
$\Big[R_Z^\beta(\tau\theta^Z_{k,b},r^Z_{(k-1)n+j,b})
\otimes R_Z^\alpha(\tau'\theta^Z_{k',b'},r^Z_{(k'-1)n+j,b'})
\Big]CZ^\epsilon$

\item[$\bullet$]
$|\beta(\phi^Z_{(k-1)n+j,b})\rangle\otimes
|\beta(\phi^Z_{(k'-1)n+j,b'})\rangle$:
$\Big[R_Z^\beta(\tau\theta^Z_{k,b},r^Z_{(k-1)n+j,b})
\otimes R_Z^\beta(\tau'\theta^Z_{k',b'},r^Z_{(k'-1)n+j,b'})
\Big]CZ^\epsilon$

\item[$\bullet$]
$|\beta(\phi^Z_{(k-1)n+j,b})\rangle\otimes
|\gamma\rangle$:
$R_Z^\beta(\tau\theta^Z_{k,b},r^Z_{(k-1)n+j,b})
\otimes Z
$

\item[$\bullet$]
$|\gamma\rangle\otimes
|\alpha(\phi^Z_{(k'-1)n+j,b'})\rangle$:
$Z\otimes 
R_Z^\alpha(\tau'\theta^Z_{k',b'},r^Z_{(k'-1)n+j,b'})
$

\item[$\bullet$]
$|\gamma\rangle\otimes
|\beta(\phi^Z_{(k'-1)n+j,b'})\rangle$:
$Z\otimes 
R_Z^\beta(\tau'\theta^Z_{k',b'},r^Z_{(k'-1)n+j,b'})
$

\item[$\bullet$]
$|\gamma\rangle\otimes
|\gamma\rangle$:
$Z\otimes Z$

\end{itemize}

\item[\bf (IV)]
If the $Z$-rotation by
$\theta^Z_{k,b}$ is implemented in the preivous step, then Alice sets
$\tau=0$.
If the $z$-rotation by
$\theta^Z_{k',b'}$
is implemented in the previous step, then Alice sets
$\tau'=0$.
If the CZ is implemented in the previous step, then Alice sets
$\epsilon=0$. 
\item[\bf (V)]
Bob does the measurement $\{|1\rangle,|2\rangle,|3\rangle\}$
on the next plain qutrit if any.
\end{itemize}
\end{algorithm}




\subsection{Ground state computing}
In the single-server protocol, it is not easy for
Bob to enjoy the energy-gap protection,
since Bob cannot prepare any natural parent Hamiltonian without knowing
$\{\xi_{a,b}^{Z/X}\}$.
This is understood as follows.

Let us consider
the span
\begin{eqnarray*}
\Big\{~
|RAKLT_b^{2N,L,R}
(\{\xi_{a,b}^{Z/X}\})
\rangle~
\Big|~
\xi_{a,b}^{Z/X}\in{\mathcal A},~a=1,...,N~
\Big\}
\end{eqnarray*}
of all rotated AKLT states. 
As is shown below,
the dimension of the span
is $2^{2N}$.
Therefore, if Bob does not know $\{\xi_{a,b}^{Z/X}\}$, he must
prepare an unnatural Hamiltonian with exponentially-degenerated ground states.

Let us show that the dimension of the span is $2^{2N}$.
Let 
\begin{eqnarray*}
|\psi_{2N}\rangle=|\phi_0\rangle\otimes
\Big(
\bigotimes_{i=1}^{2N} 
|\phi_i\rangle\Big)
\otimes
|\phi_{2N+1}\rangle,
\end{eqnarray*}
where 
$|\phi_0\rangle$
and
$|\phi_{2N+1}\rangle$
are qubit states,
and $|\phi_i\rangle$ ($i=1,...,2N$)
are qutrit states.
Let 
\begin{eqnarray*}
\mathbf{U}_{\vec{\xi}}=I_2\otimes
{\mathcal U}_b(\{\xi_{a,b}^{Z/X}\})\otimes I_2
\end{eqnarray*}
be a grobal unitary operator,
where $I_2$ is the identity operator on a single qubit.
Let $E$ is a global unitary operator which works as
\begin{eqnarray*}
E|\psi_{2N}\rangle=|AKLT^{2N,L,R}\rangle.
\end{eqnarray*}
From Lemma below and the fact that $E^\dagger$ is unitary,
\begin{eqnarray*}
\mbox{dim}\Big(\mbox{span}\Big\{\mathbf{U}_{\vec{\xi}} E|\psi_{2N}\rangle \Big\}_{\vec{\xi}}\Big)&=& 
\mbox{dim}\Big(\mbox{span}\Big\{(\mathbf{U}_{\vec{\xi}} E )^\dagger |\psi_{2N}\rangle \Big\}_{\vec{\xi}}\Big)\\
&=&\mbox{dim}\Big(\mbox{span}\Big\{E^\dagger\mathbf{U}_{\vec{\xi}}^\dagger|\psi_{2N}\rangle \Big\}_{\vec{\xi}}\Big)\\
&=&\mbox{dim}\Big(\mbox{span}\Big\{\mathbf{U}_{\vec{\xi}}^\dagger|\psi_{2N}\rangle \Big\}_{\vec{\xi}}\Big)\\
&=&\mbox{dim}\Big(\mbox{span}\Big\{\mathbf{U}_{\vec{\xi}}|\psi_{2N}\rangle \Big\}_{\vec{\xi}}\Big)\\
&=& 2^{2N}.
\end{eqnarray*}

{\bf Lemma}: Let $\{V_1, \ldots V_r\}$ be a set of $r$ operators, and let $|\phi\rangle$ be a state in their domain.
Then 
\begin{eqnarray*}
\mbox{dim}\Big(\mbox{span}\Big\{V_i|\phi\rangle\Big\}_i\Big)=\mbox{dim}\Big(\mbox{span}\Big\{
V_i^\dagger |\phi\rangle\Big\}_i\Big).
\end{eqnarray*}

{\bf Proof}:
Recall that $\mbox{dim}\Big(\mbox{span}\Big\{V_i |\phi\rangle\Big\}_i\Big)$ 
is equal to the rank of the Gram matrix of the set of vectors 
$\{V_i |\phi\rangle\}_i$.
Also note that if $G_A$ is the Gram matrix of the set of vectors  
$\{V_i |\phi\rangle\}_i$ and $G_B$ is the Gram matrix of the set of vectors $\{V_i^\dagger |\phi\rangle\}_i$, then 
$G_A = G_B^*$.
Finally, let us remind that $\mbox{rank}(A) = \mbox{rank} (A^*)$ for all matrices $A$.


\section{Blindness of the single-server protocol}
\label{Sec:single_proof}
In this section, we show the blindness of the single-server protocol (composed of Protocols 2, 3, 4). Informally speaking, a protocol is defined to be blind if Bob, given all the classical and quantum information during the protocol, cannot learn anything about the Alice's actual computational angles, input and the output \cite{blindcluster}. In the original paper for the blind quantum computation over the cluster states~\cite{blindcluster} blindness is formally defined in terms of the independence of classical and quantum states of Bob from Alice's secret. Here we adapt the definition to our setting but we omit a formal proof of the equivalences between the two definitions.
\begin{defn}
A single-server protocol is blind if 
\begin{itemize}
\item[(S1)]
The conditional probability distribution of  
Alice's nontrivial computational angles, 
given all the classical information Bob can obtain during the protocol, 
and given the measurement results of any POVMs which Bob may perform
on his system at any stage of the protocol,
is uniform,
\end{itemize}
and
\begin{itemize}
\item[(S2)]
The register state in the correlation space is one-time padded
to Bob.
\end{itemize}
\end{defn}

In order to show (S1), we have to show three lemmas.

In the following we define $\Phi,\Theta,\Xi$ and 
$R$ to be independently and uniformly distributed random variables, corresponding to the angles sent by Alice to Bob, Alice's secret computational angle, random prerotation and a hidden binary parameter, respectively.
From the construction of the protocol, the following relation is satisfied:
 \begin{eqnarray*}
\Phi=\Theta+\Xi+R\pi~~~(\mbox{mod}~2\pi).
\end{eqnarray*}
We denote by $\rho_\Xi$ the state that Alice sends to Bob parametrized by $\Xi$.
The most general method Bob may resort to in order to learn Alice's secret computational angles is described by a POVM measurement $\{\Pi_j\}_{j=1}^{m}$  on $\rho_\Xi$. This POVM can depend on all classical messages received from Alice. 
Let $O\in\{1,...,m\}$ be the random variable corresponding to the result
of the POVM measurement. Bob's knowledge about Alice's secret angles is given by the conditional probability distribution
of $\Theta=\theta$ given
 $O=j$ and  $\Phi=\phi$: 
\begin{eqnarray*}
P(\Theta=\theta|O=j,\Phi=\phi).
\end{eqnarray*}
\begin{lemma}\label{l1}

If $\rho_\Xi$ is a Dango state $|D_{Z/X}(\Xi)\rangle \langle D_{Z/X}(\Xi) |$, 
then
$
P(\Theta=\theta|O=j,\Phi=\phi)=\frac{1}{8}
$
for any 
$\theta,
\phi\in{\mathcal A}$,
$j\in\{1,...,m\}$,
and POVM on $\rho_{\Xi}$.
\end{lemma}
{\bf Proof:}
From Bayes' theorem, we have
\begin{eqnarray*}
P(\Theta=\theta|O=j,\Phi=\phi)
&=&\frac{P(O=j|\Theta=\theta,\Phi=\phi)
P(\Theta=\theta,\Phi=\phi)}
{P(O=j,\Phi=\phi)}\\
&=&\frac{P(O=j|\Theta=\theta,\Phi=\phi)
P(\Theta=\theta)P(\Phi=\phi)}
{P(O=j|\Phi=\phi)P(\Phi=\phi)}\\
&=&\frac{1}{8}\frac{\mbox{Tr}\big[
\Pi_j\frac{1}{2}
\sum_r\rho_{\phi-\theta-r\pi}\big]
}
{\mbox{Tr}\big[
\Pi_j\frac{1}{8}\frac{1}{2}\sum_{\theta,r}\rho_{\phi-\theta-r\pi}\big]}.
\end{eqnarray*}
If $\rho_\Xi$ is a Dango state $|D_{Z/X}(\Xi)\rangle \langle D_{Z/X}(\Xi) |$,
we obtain
\begin{eqnarray*}
\frac{1}{2}\sum_r\rho_{\phi-\theta-r\pi}
=\frac{1}{2}\frac{1}{8}\sum_{\theta,r}\rho_{\phi-\theta-r\pi}
\end{eqnarray*}
for any $\phi,\theta\in[0,2\pi]$,
and hence $P(\Theta=\theta|O=j,\Phi=\phi)=1/8$.
The above equation is valid since
\begin{eqnarray*}
|D_Z(\xi)\rangle\langle D_Z(\xi)|
=
\big[I\otimes(I-|\eta_1\rangle\langle\eta_1|)\otimes I\big]
\frac{1}{4}
\left(
\begin{array}{cccc}
1&e^{-i\xi}&1&1\\
e^{i\xi}&1&e^{i\xi}&e^{i\xi}\\
1&e^{-i\xi}&1&1\\
1&e^{-i\xi}&1&1
\end{array}
\right)
\oplus{\bf0}_{12}
\big[I\otimes(I-|\eta_1\rangle\langle\eta_1|)\otimes I\big],
\end{eqnarray*}
where ${\bf 0}_{12}$ is the $12\times 12$ zero matrix
and the $4\times 4$ matrix 
is in the basis 
$\{|0000\rangle,|0011\rangle,|1100\rangle,|1111\rangle\}$,
and there exists a unitary which maps 
$(I\otimes T_Z(\xi)\otimes I)|\eta_1\rangle\otimes|\eta_1\rangle$
to
$(I\otimes T_X(\xi)\otimes I)|\eta_1\rangle\otimes|\eta_1\rangle$.
$\blacksquare$


\begin{lemma}\label{l2}

Consider a collection of $L$ states $\{\rho_{\Xi_l}\}_l$ such that for each $l=1,\ldots, L$ we have
\begin{eqnarray*}
P\Big(\Theta_l=\theta_l
\Big|O^l=j,\Phi_l=\phi_l
\Big)=\frac{1}{8}
\end{eqnarray*}
where $\Phi_l$, $\Theta_l$, $\Xi_l$ and $R_l$  are defined as before.
Also, $O^l$ are the random variables corresponding to (arbitrary) POVMs performed on individual systems. 
Then for any global POVM performed on the entire collection of $L$ states  we have 

\begin{eqnarray*}
P\Big(\Theta=(\theta_1,...,\theta_L)
\Big|O=j,\Phi=(\phi_1,...,\phi_L)
\Big)=\frac{1}{8^L}
\end{eqnarray*}
where $O$ is the random variable corresponding to the outcome of the global POVM.
\end{lemma}
{\bf Proof}:
Similar to the previous proof,
from Bayes' theorem, we have
\begin{eqnarray*}
&&P\Big(\Theta=(\theta_1,...,\theta_L)
\Big|O=j,\Phi=(\phi_1,...,\phi_L)
\Big)\\
&=&
\frac{P\Big(O=j\Big|\Theta=(\theta_1,...,\theta_L),\Phi=(\phi_1,...,\phi_L)\Big)
P\Big(\Theta=(\theta_1,...,\theta_L),\Phi=(\phi_1,...,\phi_L)\Big)}
{P\Big(O=j,\Phi=(\phi_1,...,\phi_L)\Big)}\\
&=&
\frac{P\Big(O=j\Big|\Theta=(\theta_1,...,\theta_L),\Phi=(\phi_1,...,\phi_L)\Big)
P\Big(\Theta=(\theta_1,...,\theta_L)\Big)
P\Big(\Phi=(\phi_1,...,\phi_L)
\Big)}
{P\Big(O=j\Big|\Phi=(\phi_1,...,\phi_L)\Big)
P\Big(\Phi=(\phi_1,...,\phi_L)\Big)}\\
&=&
\frac{1}{8^L}
\frac{\mbox{Tr}\Big[\Pi_j\bigotimes_{i=1}^L
\frac{1}{2}\sum_{r_i}\rho_{\phi_i-\theta_i-r_i\pi}
\Big]}
{\mbox{Tr}
\Big[\Pi_j
\bigotimes_{i=1}^L
\frac{1}{8}\frac{1}{2}\sum_{\theta_i,r_i}\rho_{\phi_i-\theta_i-r_i\pi}
\Big]}.
\end{eqnarray*}

Let us define two local operators acting on $l$th state $\rho_{\Xi_l}$ by
\begin{eqnarray*}
\Pi^{l}_{j}
&\equiv&
\mbox{Tr}_{1,...,l-1,l+1,...,L}
\Big[
\Pi_j\bigotimes_{i=1}^{l-1}
\frac{1}{2}(\sum_{r_i}\rho_{\phi_i-\theta_i-r_i\pi})  
\bigotimes_{i=l+1}^{L} 
\frac{1}{2}(\sum_{r_i}\rho_{\phi_i-\theta_i-r_i\pi})
\Big],\\
\tilde{\Pi}^{l}_{j}
&\equiv&
\mbox{Tr}_{1,...,l-1,l+1,...,L}
\Big[\Pi_j
\bigotimes_{i=1}^{l-1}\frac{1}{8}\frac{1}{2}(\sum_{\theta_i,r_i}
\rho_{\phi_i-\theta_i-r_i\pi}) 
\bigotimes_{i=l+1}^{L}\frac{1}{8}\frac{1}{2}(\sum_{\theta_i,r_i}
\rho_{\phi_i-\theta_i-r_i\pi})
\Big].
\end{eqnarray*}
The partial trace is a CPTP superoperator, 
hence the above operators are
non-negative operators,
and since
\begin{eqnarray*}
\sum_{j=1}^m
\Pi^{l}_{j}
&=&I,\\
\sum_{j=1}^m
\tilde{\Pi}^{l}_{j}
&=&I,
\end{eqnarray*}
hence
$\{\Pi^l_j\}_{j=1}^m$ and
$\{\tilde{\Pi}^l_j\}_{j=1}^m$ 
are local POVMs on $l$th state.

Let $O^l$ and $\tilde{O}^l$ 
be the random variables which correspond to the results of the POVMs
$\{\Pi^l_j\}_{j=1}^m$ and $\{\tilde{\Pi}^l_j\}_{j=1}^m$,
respectively.
Then, we have
\begin{eqnarray*}
P\Big(
\Theta=(\theta_1,...,\theta_L)\Big|
O=j,\Phi=(\phi_1,...,\phi_L)
\Big)
&=&
\frac{1}{8^L}
\frac{
\mbox{Tr}_l
\Big[
\Pi^{l}_{j}
\frac{1}{2}\sum_{r_l}\rho_{\phi_l-\theta_l-r_l\pi}
\Big]}
{
\mbox{Tr}_l
\Big[
\tilde{\Pi}^{l}_{j}
\frac{1}{8}\frac{1}{2}\sum_{\theta_l,r_l}\rho_{\phi_l-\theta_l-r_l\pi}
\Big]}\\
&=&
\frac{1}{8^L}
\frac{P(O^l=j|\Phi_l=\phi_l,\Theta_l=\theta_l)}
{P(\tilde{O}^l=j|\Phi_l=\phi_l)}\\
&=&
\frac{1}{8^L}
\frac{P(O^l=j|\Phi_l=\phi_l,\Theta_l=\theta_l)}
{P(O^l=j|\Phi_l=\phi_l)}
\frac
{P(O^l=j|\Phi_l=\phi_l)}
{P(\tilde{O}^l=j|\Phi_l=\phi_l)}\\
&=&
\frac{1}{8^{L-1}}
\frac{P(O^l=j|\Phi_l=\phi_l,\Theta_l=\theta_l)P(\Phi_l=\phi_l)P(\Theta_l=\theta_l)}
{P(O^l=j|\Phi_l=\phi_l)P(\Phi_l=\phi_l)}
\frac
{P(O^l=j|\Phi_l=\phi_l)}
{P(\tilde{O}^l=j|\Phi_l=\phi_l)}\\
&=&
\frac{1}{8^{L-1}}
\frac{P(O^l=j|\Phi_l=\phi_l,\Theta_l=\theta_l)P(\Phi_l=\phi_l,\Theta_l=\theta_l)}
{P(O^l=j,\Phi_l=\phi_l)}
\frac
{P(O^l=j|\Phi_l=\phi_l)}
{P(\tilde{O}^l=j|\Phi_l=\phi_l)}\\
&=&
\frac{1}{8^{L-1}}
P(\Theta_l=\theta_l|\Phi_l=\phi_l,O^l=j)
\frac
{P(O^l=j|\Phi_l=\phi_l)}
{P(\tilde{O}^l=j|\Phi_l=\phi_l)}\\
&=&
\frac{1}{8^L}
\frac
{P(O^l=j|\Phi_l=\phi_l)}
{P(\tilde{O}^l=j|\Phi_l=\phi_l)}.
\end{eqnarray*} 
Note that
$P(O^l=j|\Phi_l=\phi_l)$ and $P(\tilde{O}^l=j|\Phi_l=\phi_l)$
are independent of $\theta_l$,
and hence
\begin{eqnarray*}
P\Big(\Theta=
(\theta_1,...,\theta_L)\Big|
O=j,
\Phi=(\phi_1,...,\phi_L)\Big)
\end{eqnarray*}
is also independent of $\theta_l$.
The same result holds for any $l=1,...,L$, hence the
proof is completed.
$\blacksquare$

Recall that a single $X/Z$ Dango chain is used for the implementation of a fixed $X/Z$ rotation. The following lemma shows this information does not help Bob to learn about Alice's secret. 
\begin{lemma}\label{l3}
Under the assumptions of Lemma \ref{l2},
assume that
$\Theta$ takes values with a non-zero probability only in a subset $K$,
i.e. $\Theta\in K\subset{\mathcal A}^{\times L}$.
Then
\begin{eqnarray*}
P\Big(\Theta=\theta
\Big|O=j,\Phi=(\phi_1,...,\phi_L),
\Theta\in K
\Big)=\frac{1}{|K|}
\end{eqnarray*}
for any 
$\theta\in K$,
$(\phi_1,...,\phi_L)\in{\mathcal A}^{\times L}$,
$j\in\{1,...,m\}$,
and POVM on $\bigotimes_{i=1}^L\rho_{\Xi_i}$.
\end{lemma}
{\bf Proof}:
Similar to the previous proofs, we have
\begin{eqnarray*}
P\Big(\Theta=\theta
\Big|O=j,\Phi=(\phi_1,...,\phi_L),
\Theta\in K
\Big)
&=&
\frac{P\Big(\Theta=\theta,
\Theta\in K
\Big|O=j,\Phi=(\phi_1,...,\phi_L)
\Big)}
{P\Big(\Theta\in K\Big)}\\
&=&
\frac{P\Big(\Theta=\theta
\Big|O=j,\Phi=(\phi_1,...,\phi_L)
\Big)}
{P\Big(\Theta\in K\Big)}\\
&=&
\frac{1/8^L}
{|K|/8^L}.
\end{eqnarray*}
$\blacksquare$

\begin{theorem}\label{t1}
The single-server protocol satisfies (S1).
\end{theorem}
{\bf Proof}:
Bob receives $2NM$ Dango states, and therefore there are
$2NM$ secret angles $\big\{\xi_{a,b}^{Z/X}\big\}_{(a,b)=(1,1)}^{(N,M)}$.
Bob also receives $2NM$ angles  
$\big\{\phi_{a,b}^{Z/X}\big\}_{(a,b)=(1,1)}^{(N,M)}$ from Alice.
Let
$\Theta,\Phi\in{\mathcal A}^{\times 2NM}$
be random variables, and $O\in\{1,...,m\}$
be the random variable which corresponds to the result
of the POVM measurement which Bob performs
on his system.
Since Bob knows that Alice tries the same rotation many times in a single $Z/X$-AKLT subsystem
until she succeeds, and that after the success of the rotation 
Alice implements
the trivial identity operation on the rest of qutrits in
the $Z/X$-AKLT subsystem, 
Bob can assume that $\Theta$ takes values only in a subset $K$:
$\Theta\in K\subset{\mathcal A}^{\times 2NM}$,
where 
$|K|=8^{2NM/n}$.

From Lemma 1, 2, and 3, we have the following equality $\forall \theta\in K$,
$\forall \phi_{a,b}^{Z/X}\in{\mathcal A}$,
$(a=1,...,N)$,
$(b=1,...,M)$,
$j\in\{1,...,m\}$,
and for any POVM
\begin{eqnarray*}
P\Big(
\Theta=\theta
\Big|
O=j, 
\Phi=
\big\{
\phi_{a,b}^{Z/X}
\big\}_{(a,b)=(1,1)}^{(N,M)},
\Theta\in K
\Big)=\frac{1}{|K|}
\end{eqnarray*} 
$\blacksquare$

\begin{theorem}\label{t1}
The single-server protocol satisfies (S2).
\end{theorem}
{\bf Proof}:
It is easy to see 
\begin{itemize}
\item
When $R^\alpha_Z(\theta^Z_{k,b},r^Z_{(k-1)n+j,b})$
or $R^\beta_Z(\theta^Z_{k,b},r^Z_{(k-1)n+j,b})$
is implemented in the correlation space, 
the byproduct
$XZ^{r^Z_{(k-1)n+j,b}}$ or
$XZ^{r^Z_{(k-1)n+j,b}+1}$ 
occurs, respectively.
If Bob has no information about the
value of $r^Z_{(k-1)n+j,b}$, he cannot know
whether the byproduct $Z$ appears or not.
\item
When $R^\alpha_X(\theta^X_{k,b},r^X_{(k-1)n+j,b})$
or $R^\beta_X(\theta^X_{k,b},r^X_{(k-1)n+j,b})$
is implemented in the correlation space, 
the byproduct
$X^{r^X_{(k-1)n+j,b}+1}Z$ or
$X^{r^X_{(k-1)n+j,b}}Z$ 
occurs, respectively.
If Bob has no information about the value of
$r^X_{(k-1)n+j,b}$, he cannot know
whether the byproduct $X$ appears or not.
\end{itemize}
Note that we assume Alice's computation is implemented via a regular
structure hence it contains both $X$ and $Z$ rotations.
Therefore both byproducts of Pauli $X$ and $Z$ 
operators will appear leading to the full one-time padding
of the computation in the correlation space.
In fact, we can show that
Bob cannot have any information about the values
of
$\{r^{Z/X}_{(k-1)n+j,b}\}$ 
by showing similar proofs
as those for $\{\theta_{k,b}^{Z/X}\}$.  
However,
it is easy to consider that $\theta_{k,b}^{Z/X}$ takes values only
$0$ or $\pi$ in the above proofs leading
to the exchange of the role of
$\{r^{Z/X}_{(k-1)n+j,b}\}$ 
and
$\{\theta_{k,b}^{Z/X}\}$.  

\section{Double-server protocol}
\label{Sec:double}

In this section, we will explain the double-server protocol (Fig.~\ref{double_plain}). There are two advantages behind the new protocol: Alice could be completely classical and more importantly the resource state preparation and computation could be done more robustly using the energy-gap protection. To achieve these new features while keeping the security requirement intact, it is assumed that the two servers, Bob1 and Bob2, share many Bell pairs but have no classical or quantum channel between them. As we will discuss later it is an interesting open question (both from the practical and theoretical perspective) whether this assumption could be relaxed.

In the double-server protocol, Bob1 first creates AKLT resource states (without any random rotation), hence the preparation and storage of the state could be performed using ground state energy-gap protection as described in details in \cite{MiyakeAKLT}. Next, depending on Alice's desired gates,
Bob1 adiabatically turns off the interaction between some particles and the rest of
particles in his resource state, and teleports these particles to Bob2 by consuming Bell pairs. Bob1 sends Alice the result of the measurement in the teleportation through the classical channel. Note that due to the lack of any communication (classical or quantum) channels between Bob1 and Bob2, the teleportation procedure from Bobs' point of view can be seen as a usage of a totally mixed channel where only Alice knows how to correct the output of the channel.

Next, Alice calculates the angle in which particles should be measured by using her classical computer, and sends Bob2 the angle which is the sum of the calculated angle plus the compensation for the byproduct and a random angle to hide the actual angle of the computation. Bob2 performs the measurement in that angle and sends the result of the measurement to Alice. Next, Alice sends the previous random angle to Bob1 and he does the single-qubit rotation which compensates the added random angle.  Bob1 and Bob2 repeat this two-way  classical communication with Alice until they finish the computation.

\begin{figure}[htbp]
\begin{center}
\includegraphics[width=0.6\textwidth]{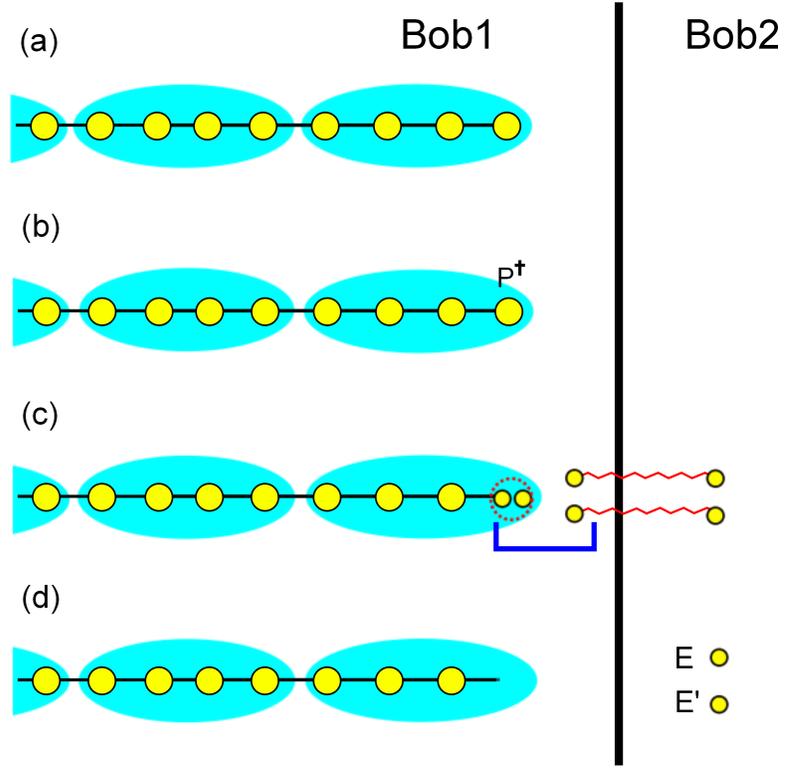}
\end{center}
\caption{(Color online.) 
(a): A chain $|AKLT^{N,L,R}\rangle$. Each AKLT subsystem is specified with a blue circle.
In this example, $n=4$.
(b): Bob1 adiabatically turns off the interaction
between a qutrit
and others,
and
applies $P^\dagger$ to the qutrit
in order to convert this qutrit into a pair of two qubits.
(c): Bob1 teleports thus created two qubits 
to Bob2 by consuming two Bell pairs. 
(d): Because of the teleportation,
Pauli errors $E$ and $E'$ occur.
} 
\label{spin1_teleportation_plain}
\end{figure}
 
We define a $(k,b)$th AKLT subsystem ($k=1,...,N/n$, $b=1,...,M$) to be the collection of $n$ qutrits of the $b$th AKLT chain with column index $(k-1)n+1,(k-1)n+2,...,(k-1)n+n$ (Fig.~\ref{spin1_teleportation_plain} (a)). A single-qubit rotation is implemented in a single AKLT subsystem. Let us assume that Alice wants to perform the single-qubit
$Z$-rotation $\exp\Big[\frac{iZ}{2}\theta^Z_{k,b}\Big]$ with  $\theta^Z_{k,b}\in{\mathcal A}$ using $(k,b)$th AKLT subsystem (Protocol 5).
The protocol for implementing an arbitrary $X$-rotation is similar and is given in the Appendix \ref{App:Xrot}. Finally, in order to perform blind $CZ$ gates, similar to the single-sever protocol, Bob1 periodically implements $CZ$ gates.
In order to keep the register state in the ground space,
the interactions are adiabatically turned off before each $CZ$ gate.
Unwanted $CZ$ gates are canceled in the same way as that in
the single-server protocol.

\begin{algorithm}
\caption{\bf 5: Double-server Blind $Z$ rotation}
\label{prot:2Zrot}

Initially the flag parameter (known to only Alice) is set $\tau=1$. Alice sets her secret parameter $\epsilon^Z_{k,b}=0$ and also chooses  
random numbers $\delta^Z_{k,b}\in{\mathcal A}$. 
Alice sends Bob1 parameter values $N$, $M$ and $n<N$. 
Bob1 creates $M$ AKLT chains $|AKLT^{N,L,R}_b\rangle$, where $b=1,...,M$ (see Equation \ref{eq:AKLT}) 
of $N$ qutrits arranged in an array of $N$ columns and $M$ rows. 
For $j=1 \cdots n/2$ Alice, Bob1, and Bob2 repeat (I)-(VI).
\begin{itemize}
\item[\bf (I)]
Bob1 adiabatically turns off the interaction which acts on the
$j$th qutrit of $(k,b)$th AKLT subsystem
and applies $P^\dagger$ to the isolated qutrit
in order to convert the qutrit into the pair of two qubits
(Fig.~\ref{spin1_teleportation_plain} (b)).
(The application of $P^\dagger$ can be done deterministically.
For the physical implementation of $P^\dagger$,
see Appendix~\ref{App:Pinverse}.) 

\item[\bf (II)]
Bob1 teleports the created two qubits
to Bob2 by consuming two Bell
pairs (Fig.~\ref{spin1_teleportation_plain} (c)).
These two teleported qubits are affected by a two-qubit Pauli error 
$E\otimes E'\in\{I,X,Z,XZ\}\otimes\{I,X,Z,XZ\}$ (Fig.~\ref{spin1_teleportation_plain} (d)). Bob1 sends Alice the result of the measurement in the teleportation and hence only Alice and Bob1 knows which error appears.

\item[\bf (III)]
Bob2 applies the filtering operation
$\{I-|\eta_1\rangle\langle\eta_1|,|\eta_1\rangle\langle\eta_1|\}$
to the received two qubits, 
and sends the result to Alice.
\begin{itemize}
\item[$\bullet$]
If
the Pauli error is
$I\otimes I$,
$X\otimes X$, $Z\otimes Z$, or $XZ\otimes XZ$,
the probability of realizing $|\eta_1\rangle\langle\eta_1|$ is 0.

\item[$\bullet$]
If the Pauli error is
$I\otimes Z$, $X\otimes XZ$, $Z\otimes I$, or $XZ\otimes X$,
$|\eta_1\rangle\langle\eta_1|$ is realized with
the probability $1/3$. If $|\eta_1\rangle\langle\eta_1|$
is realized, $Z$ is implemented
in the correlation space.

\item[$\bullet$]
If the Pauli error is
$I\otimes X$, $X\otimes I$, $Z\otimes XZ$, or $XZ\otimes Z$,
$|\eta_1\rangle\langle\eta_1|$ is realized with
the probability $1/3$. If $|\eta_1\rangle\langle\eta_1|$ is
realized, 
$X$
is implemented in the correlation space.

\item[$\bullet$]
If the Pauli error is
$I\otimes XZ$, $X\otimes Z$, $Z\otimes X$, or $XZ\otimes I$,
$|\eta_1\rangle\langle\eta_1|$ is realized with
the probability $1/3$. If $|\eta_1\rangle\langle\eta_1|$
is realized, 
$XZ$
is implemented in the correlation space.

\end{itemize}
If $|\eta_1\rangle\langle\eta_1|$ is realized, 
skip steps ({\bf IV}), ({\bf V}) and ({\bf VI}). 
If $I-|\eta_1\rangle\langle\eta_1|$ is realized,
Bob2 further applies the PEPS operation $P$ on the two qubits.
This PEPS operation is done deterministically, because the two qubits
are already projected by $I-|\eta_1\rangle\langle\eta_1|$.

\item[\bf (IV)]
Alice sends the angle $\phi^Z_{(k-1)n+j,b}$
to Bob2. This angle is determined according to the following rule:
\begin{itemize}
\item[$\bullet$]
If the Pauli error is 
$I\otimes I$,
$I\otimes Z$,
$Z\otimes I$,
or $Z\otimes Z$,
\begin{eqnarray*}
\phi^Z_{(k-1)n+j,b}=\tau\theta^Z_{k,b}+\tau\delta^Z_{k,b}+\xi^Z_{(k-1)n+j,b}+
r^Z_{(k-1)n+j,b}\pi~~~(\mbox{mod}~2\pi),
\end{eqnarray*}
where $\xi^Z_{(k-1)n+j,b}\in{\mathcal A}$ and 
$r^Z_{(k-1)n+j,b}\in\{0,1\}$ are random numbers chosen by Alice,
and signs of $\theta^Z_{k,b}$ and $\delta^Z_{k,b}$ should be changed if there is the
byproduct $X$ before this step.

\item[$\bullet$]
If the Pauli error is 
$X\otimes X$,
$X\otimes XZ$,
$XZ\otimes X$,
or $XZ\otimes XZ$,
\begin{eqnarray*}
\phi^Z_{(k-1)n+j,b}=-\tau\theta^Z_{k,b}-\tau\delta^Z_{k,b}-\xi^Z_{(k-1)n+j,b}
+r^Z_{(k-1)n+j,b}\pi~~~(\mbox{mod}~2\pi),
\end{eqnarray*}
where $\xi^Z_{(k-1)n+j,b}\in{\mathcal A}$ and 
$r^Z_{(k-1)n+j,b}\in\{0,1\}$ are random numbers chosen by Alice,
and signs of $\theta^Z_{k,b}$ and $\delta^Z_{k,b}$ should be changed if there is the
byproduct $X$ before this step.

\item[$\bullet$]
If the Pauli error is
$I\otimes X$,
$I\otimes XZ$,
$X\otimes I$,
$X\otimes Z$,
$Z\otimes X$,
$Z\otimes XZ$,
$XZ\otimes I$,
or
$XZ\otimes Z$,
\begin{eqnarray*}
\phi^Z_{(k-1)n+j,b}=\xi^Z_{(k-1)n+j,b}
\end{eqnarray*}
where
$\xi^Z_{(k-1)n+j,b}\in{\mathcal A}$
is a random number chosen by Alice.
\end{itemize}

\item[\bf (V)]
Bob2 does the measurement $\mathcal{M}(\phi^Z_{(k-1)n+j,b})$ (similar to
Protocol 2), 
and sends the result to Alice.
By this measurement, following operations are implemented 
in the correlation space (see also Appendix~\ref{App:tele2}):
\begin{itemize}

\item[$\bullet$]
If the Pauli error is $I\otimes I$ or $Z\otimes Z$,
$|\alpha(\phi^Z_{(k-1)n+j,b})\rangle$,
$|\beta(\phi^Z_{(k-1)n+j,b})\rangle$,
or 
$|\gamma\rangle$
occurs with the probability $1/3$ respectively.
If $|\alpha(\phi^Z_{(k-1)n+j,b})\rangle$ is realized,
\begin{eqnarray*}
XZ^{r^Z_{(k-1)n+j,b}}\exp\Big[\frac{iZ}{2}\Big(\tau\theta^Z_{k,b}+
\tau\delta^Z_{k,b}
+\xi^Z_{(k-1)n+j,b}\Big)\Big]
\end{eqnarray*}
is implemented.
If $|\beta(\phi^Z_{(k-1)n+j,b})\rangle$ is realized,
\begin{eqnarray*}
XZ^{r^Z_{(k-1)n+j,b}+1}
\exp\Big[\frac{iZ}{2}\Big(\tau\theta^Z_{k,b}+\tau\delta^Z_{k,b}+
\xi^Z_{(k-1)+j,b}\Big)\Big] 
\end{eqnarray*}
is implemented.
If $|\gamma\rangle$ is realized,
$Z$ is implemented.
\end{itemize}
\end{itemize}
\end{algorithm}

\begin{algorithm}
\caption{\bf 5:  --- Continued}
\begin{itemize}
\item[\bf (V)] ---Continued
\begin{itemize}
\item[$\bullet$]
If the Pauli error is $I\otimes Z$ or $Z\otimes I$,
$|\alpha(\phi^Z_{(k-1)n+j,b})\rangle$ or
$|\beta(\phi^Z_{(k-1)n+j,b})\rangle$
occurs with the probability $1/2$ respectively.
If $|\alpha(\phi^Z_{(k-1)n+j,b})\rangle$ is realized,
\begin{eqnarray*}
XZ^{r^Z_{(k-1)n+j,b}+1}
\exp\Big[\frac{iZ}{2}\Big(
\tau\theta^Z_{k,b}+\tau\delta^Z_{k,b}+\xi^Z_{(k-1)n+j,b}\Big)\Big] 
\end{eqnarray*}
is implemented.
If $|\beta(\phi^Z_{(k-1)n+j,b})\rangle$ is realized,
\begin{eqnarray*}
XZ^{r^Z_{(k-1)n+j,b}}
\exp\Big[\frac{iZ}{2}\Big(
\tau\theta^Z_{k,b}+\tau\delta^Z_{k,b}+\xi^Z_{(k-1)n+j,b}\Big)\Big]
\end{eqnarray*}
is implemented.

\item[$\bullet$]
If the Pauli error is $X\otimes X$ or $XZ\otimes XZ$,
$|\alpha(\phi^Z_{(k-1)n+j,b})\rangle$, 
$|\beta(\phi^Z_{(k-1)n+j,b})\rangle$, or
$|\gamma\rangle$ occurs with the probability $1/3$ respectively.
If
$|\alpha(\phi^Z_{(k-1)n+j,b})\rangle$ is realized,
\begin{eqnarray*}
XZ^{r^Z_{(k-1)n+j,b}}
\exp\Big[\frac{iZ}{2}\Big(
\tau\theta^Z_{k,b}+\tau\delta^Z_{k,b}+\xi^Z_{(k-1)n+j,b}\Big)
\Big]
\end{eqnarray*}
is implemented.
If
$|\beta(\phi^Z_{(k-1)n+j,b})\rangle$ is realized,
\begin{eqnarray*}
XZ^{r^Z_{(k-1)n+j,b}+1}
\exp\Big[\frac{iZ}{2}\Big(
\tau\theta^Z_{k,b}+\tau\delta^Z_{k,b}+\xi^Z_{(k-1)n+j,b}\Big)\Big]
\end{eqnarray*}
is implemented.
If $|\gamma\rangle$ is realized,
$Z$ is implemented.

\item[$\bullet$]
If the Pauli error is $X\otimes XZ$ or $XZ\otimes X$,
$|\alpha(\phi^Z_{(k-1)n+j,b})\rangle$ or
$|\beta(\phi^Z_{(k-1)n+j,b})\rangle$
occurs with the probability $1/2$ respectively.
If
$|\alpha(\phi^Z_{(k-1)n+j,b})\rangle$ is realized,
\begin{eqnarray*}
XZ^{r^Z_{(k-1)n+j,b}+1}
\exp\Big[\frac{iZ}{2}\Big(
\tau\theta^Z_{k,b}+\tau\delta^Z_{k,b}+\xi^Z_{(k-1)n+j,b}\Big)\Big]
\end{eqnarray*}
is implemented.
If
$|\beta(\phi^Z_{(k-1)n+j,b})\rangle$ is realized,
\begin{eqnarray*}
XZ^{r^Z_{(k-1)n+j,b}}
\exp\Big[\frac{iZ}{2}\Big(
\tau\theta^Z_{k,b}+\tau\delta^Z_{k,b}+\xi^Z_{(k-1)n+j,b}\Big)
\Big]
\end{eqnarray*}
is implemented.

\item[$\bullet$]
If the Pauli error is $I\otimes X$,
$X\otimes I$, $Z\otimes XZ$,
or $XZ\otimes Z$,
$|\alpha(\phi^Z_{(k-1)n+j,b})\rangle$,  
$|\beta(\phi^Z_{(k-1)n+j,b})\rangle$, or
$|\gamma\rangle$
occurs with the probability 
$\frac{1}{2}\sin^2[\frac{1}{2}\phi^Z_{(k-1)n+j,b}]$,
$\frac{1}{2}\cos^2[\frac{1}{2}\phi^Z_{(k-1)n+j,b}]$,
or $1/2$, respectively.
If
$|\alpha(\phi^Z_{(k-1)n+j,b})\rangle$ or
$|\beta(\phi^Z_{(k-1)n+j,b})\rangle$ is realized,
$Z$ is implemented.
If $|\gamma\rangle$ is realized,
$XZ$  
is implemented.

\item[$\bullet$]
If the Pauli error is $I\otimes XZ$,
$X\otimes Z$, $Z\otimes X$,
or $XZ\otimes I$,
$|\alpha(\phi^Z_{(k-1)n+j,b})\rangle$,  
$|\beta(\phi^Z_{(k-1)n+j,b})\rangle$, or
$|\gamma\rangle$
occurs with the probability 
$\frac{1}{2}\cos^2[\frac{1}{2}\phi^Z_{(k-1)n+j,b}]$,
$\frac{1}{2}\sin^2[\frac{1}{2}\phi^Z_{(k-1)n+j,b}]$,
or $1/2$, respectively.
If
$|\alpha(\phi^Z_{(k-1)n+j,b})\rangle$ or
$|\beta(\phi^Z_{(k-1)n+j,b})\rangle$ is realized,
$Z$ is implemented.
If $|\gamma\rangle$ is realized,
$X$  
is implemented.

\end{itemize}

\item[\bf (VI)]
If the $z$-rotation 
$\exp[\frac{iZ}{2}(\tau\theta^Z_{k,b}+\tau\delta^Z_{k,b}+\xi^Z_{(k-1)n+j,b})]$
is implemented in the previous step, Alice sets $\tau=0$,
and
$\epsilon^Z_{k,b}=\epsilon^Z_{k,b}+
\xi^Z_{(k-1)n+j,b}$ $(\mbox{mod}~2\pi)$ 
(if there is no $X$ byproduct before this rotation)
or
$\epsilon^Z_{k,b}=\epsilon^Z_{k,b}-
\xi^Z_{(k-1)n+j,b}$ $(\mbox{mod}~2\pi)$
(if there is the $X$ byproduct before this rotation).

\item[\bf (VII)]
So far, 
the $z$-rotation 
\begin{eqnarray*}
G^Z_{k,b}
\exp\Big[\frac{iZ}{2}
(\epsilon^Z_{k,b}+\delta^Z_{k,b})\Big]
\exp\Big[\frac{iZ}{2}\theta^Z_{k,b}\Big]
\end{eqnarray*}
up to some Pauli byproduct $G^Z_{k,b}$ is implemented.
The probability that they fail to perform this $z$-rotation
is $(2/3)^{n/2}$,
which is small for sufficiently large $n$.
Alice asks Bob1 to correct the accumulated error. In order to do so she sends Bob1 the angle 
$\tilde{\epsilon}^Z_{k,b}=\epsilon^Z_{k,b}+\delta^Z_{k,b}$ $(\mbox{mod}~2\pi)$ 
if $G^Z_{k,b}$ contains
no $X$ byproduct, and $\tilde{\epsilon}^Z_{k,b}=-\epsilon^Z_{k,b}-\delta^Z_{k,b}$ $(\mbox{mod}~2\pi)$ if
$G^Z_{k,b}$ contains the $X$ byproduct.
Bob1 implements the rotation
$\exp\Big[-\frac{iZ}{2}\tilde{\epsilon}^Z_{k,b}\Big]$
by using the rest of the qutrits in $(k,b)$th AKLT subsystem.
The probability that Bob1 fails to perform this $z$-rotation
is $(1/3)^{n/2}$,
which is small for sufficiently large $n$.

\end{itemize}
\end{algorithm}
\section{Blindness of the double-server Protocol}
\label{Sec:double_proof}
In this section, we will show the blindness of the double-server
protocol. 

\begin{defn}
A double-sever protocol is blind if 
\begin{itemize}
\item[(D1)]
The conditional probability distribution of Alice's
nontrivial computational angles, 
given all the classical information
Bob1 can obtain during the protocol,
and given the measurement results of any POVMs which Bob1 may perform
on his system at any stage of the protocol, is uniform,
\item[(D2)]
The conditional probability distribution of Alice's nontrivial
computational angles,
given all the classical information
Bob2 can obtain during the protocol,
and given the measurement results of any POVMs which Bob2 may perform
on his system at any stage of the protocol, is uniform,
\item[(D3)]
The register state in the correlation space is one-time padded to Bob1,
\item[(D4)]
The register state in the correlation space is one-time padded to Bob2.
\end{itemize}
\end{defn}

The proof is based on following lemmas. 

\begin{lemma}\label{l4}
Bob1 cannot send any information to Bob2.
\end{lemma}

{\bf Proof}:
By the assumption, there is no channel between
Bob1 and Bob2.
Furthermore,
Bob1 cannot send any information to Bob2
via Alice either. This is due to the following facts.
First, what Bob1 sends to Alice are the ``results of measurements in
teleportations".
Second, what Alice sends to Bob2 are  
$\{\phi^{Z/X}_{(k-1)n+j,b}\}$. Third, recall that
Alice chooses the definition of each $\phi_{(k-1)n+j,b}^{Z/X}$
among
\begin{eqnarray*}
\phi^{Z/X}_{(k-1)n+j,b}&=&
\tau\theta^{Z/X}_{k,b}+\tau\delta^{Z/X}_{k,b}+\xi^{Z/X}_{(k-1)n+j,b}+r^{Z/X}_{(k-1)n+j,b}\pi~~(\mbox{mod}~2\pi),\\ 
\phi^{Z/X}_{(k-1)n+j,b}&=&
-\tau\theta^{Z/X}_{k,b}-\tau\delta^{Z/X}_{k,b}-\xi^{Z/X}_{(k-1)n+j,b}+r^{Z/X}_{(k-1)n+j,b}\pi~~(\mbox{mod}~2\pi),\\ 
\end{eqnarray*}
or
\begin{eqnarray*}
\phi^{Z/X}_{(k-1)n+j,b}=
\xi^{Z/X}_{(k-1)n+j,b}
\end{eqnarray*}
according to what Bob1 sends to Alice.
However, the value of each $\phi^{Z/X}_{(k-1)n+j,b}$ is independent of
what Bob1 sends to Alice,
since
$\xi^{Z/X}_{(k-1)n+j,b}$ 
is completely random and therefore
$\phi^{Z/X}_{(k-1)n+j,b}$ takes any value in $\mathcal A$
with equal probability, 
whichever definition 
Alice chooses.
Therefore,
\begin{eqnarray*}
P\Big(T=t\Big|\Phi=\{\phi^{Z/X}_{(k-1)n+j,b}\}\Big)
&=&
\frac{P\Big(\Phi=\{\phi^{Z/X}_{(k-1)n+j,b}\}\Big|T=t\Big)
P\Big(T=t\Big)}
{P\Big(\Phi=\{\phi^{Z/X}_{(k-1)n+j,b}\}\Big)}\\
&=&P\Big(T=t\Big),
\end{eqnarray*}
where 
$T$
is the random variable which represents teleportation results.
 
Finally, Bell pairs shared between Bob1 and Bob2 cannot 
transmit any information from Bob1 to Bob2 without sending
a classical message from Bob1 to Bob2,
since if it is possible, information is transfered faster than
light from Bob1 to Bob2. $\blacksquare$

\begin{lemma}\label{l5}
Bob2 cannot send any information to Bob1.
\end{lemma}
{\bf Proof}:
Similar to the previous lemma, Bob2 cannot
send any information to Bob1
via Alice.
This is due to the following facts.
First, what Bob2 sends to Alice are the ``results of filterings and
measurements".
Second, what Alice sends to Bob1 are $\{\tilde{\epsilon}^{Z/X}_{k,b}\}$.
Third, 
although 
$\epsilon^{Z/X}_{k,b}$
depends on 
what Bob2 sends Alice,
$\tilde{\epsilon}^{Z/X}_{k,b}$
is independent of 
what Bob2 sends Alice,
since $\delta^{Z/X}_{k,b}$
is completely random.
Therefore,
\begin{eqnarray*}
P\Big(F=f\Big|E=\{\tilde{\epsilon}^{Z/X}_{k,b}\}\Big)
&=&
\frac{P\Big(E=\{\tilde{\epsilon}^{Z/X}_{k,b}\}\Big|F=f\Big)
P\Big(F=f\Big)}
{P\Big(E=\{\tilde{\epsilon}^{Z/X}_{k,b}\}\Big)}\\
&=&P\Big(F=f\Big),
\end{eqnarray*}
where 
$F$
is the random variable which represents the results of filterings
and measurements. $\blacksquare$

\begin{theorem}
The double-server protocol satisfies (D2).
\end{theorem}
{\bf Proof}: From Lemma 4, Bob1 cannot send any information to Bob2.
Therefore, all quantum states which Bob2 receives
are completely mixed states,
and all classical information which Bob2 gains
are only $\{\phi^{Z/X}_{(k-1)n+j,b}\}$.
Since $\{\xi^{Z/X}_{(k-1)n+j,b}\}$ and $\{\delta^{Z/X}_{k,b}\}$ 
are completely random
and independent from
$\{\theta^{Z/X}_{k,b}\}$, Bob2 cannot have any information
about
$\{\theta^{Z/X}_{k,b}\}$
from
$\{\phi^{Z/X}_{(k-1)n+j,b}\}$.
$\blacksquare$


\begin{theorem}
The double-server protocol satisfies (D4).
\end{theorem}
{\bf Proof}:
It is easy to see that
\begin{itemize}
\item
When the $z$-rotation 
$\exp\Big[\frac{iZ}{2}\Big(\theta^Z_{k,b}+\delta^Z_{k,b}+\xi^Z_{(k-1)n+j,b}\Big)\Big]$
is implemented, 
the byproduct $XZ^{r^Z_{(k-1)n+j,b}}$ or $XZ^{r^Z_{(k-1)n+j,b}+1}$ occurs.
\item
When the $x$-rotation 
$\exp\Big[\frac{-iX}{2}\Big(\theta^X_{k,b}+\delta^X_{k,b}+\xi^X_{(k-1)n+j,b}\Big)\Big]$
is implemented, 
the byproduct $X^{r^X_{(k-1)n+j,b}}Z$ or $X^{r^X_{(k-1)n+j,b}+1}Z$ occurs.
\end{itemize}
Bob2 cannot gain any information about  
$\{r^{Z/X}_{(k-1)n+j,b}\}$ from 
$\{\phi^{Z/X}_{(k-1)n+j,b}\}$,
since
$\{\xi^{Z/X}_{(k-1)n+j,b}\}$ and $\{\delta^{Z/X}_{k,b}\}$ 
are completely random
and independent from
$\{r^{Z/X}_{(k-1)n+j,b}\}$.
$\blacksquare$

\begin{theorem}
The double-server protocol satisfies (D3).
\end{theorem}
{\bf Proof}:
First, since $\{\xi^{Z/X}_{(k-1)n+j,b}\}$ are completely random, 
hence $\{\epsilon^{Z/X}_{k,b}\}$
is independent from 
$\{r^{Z/X}_{(k-1)n+j,b}\}$. 
Second, although
$\tilde{\epsilon}^{Z/X}_{k,b}$
is related to 
the parity of $X$ or $Z$ in $G^{Z/X}_{k,b}$,
Bob1 cannot know these parities
from
$\{\tilde{\epsilon}^{Z/X}_{k,b}\}$
since Bob1 does not know 
$\{\delta^{Z/X}_{k,b}\}$. 
Therefore (D3) is satisfied.
$\blacksquare$

\begin{theorem}
The double-server protocol satisfies (D1).
\end{theorem}
{\bf Proof}:
From Lemma 5, Bob2 cannot send any information
to Bob1.
Therefore, the classical information which Bob1 can gain
are only
$\{\tilde{\epsilon}^{Z/X}_{k,b}\}$.
Obviously, 
$\{\tilde{\epsilon}^{Z/X}_{k,b}\}$
are independent from
$\{\theta^{Z/X}_{k,b}\}$.
Furthermore, from Theorem 5, Bob1's states are one-time padded to him,
therefore, no POVM on Bob1's states gives information to Bob1.
$\blacksquare$

\section{Discussion}
\label{Sec:disc}
In this paper we have investigated 
possibilities of blind quantum computing 
on AKLT states, which are physically-motivated resource states
for measurement-based quantum computation.
We have shown that our blind quantum computation protocols
can enjoy several advantages of AKLT states, such as
the cooling preparation of the resource state,
the energy-gap protection of the quantum computation,
and the simple and efficient preparation of the resource state
in linear optics with biphotons.

We have presented two protocols, namely the 
single-sever protocol and the double-server protocol.
In the single-server protocol,
we have shown that
blind quantum computation is possible, if Alice 
is allowed to have weak quantum technology
as in the case of Ref.~\cite{blindcluster}. 
Because AKLT resources states
exhibit many different properties from those of the cluster state,
several novel features have been introduced in our protocol compared with 
Ref.~\cite{blindcluster}. 
Furthermore, new techniques and formalism have been required
for the proof of the blindness 
of our protocol.

We have also seen that
in our single-server protocol, quantum computation 
cannot be performed in the ground space of a natural Hamiltonian, 
since if Bob does not know pre-rotated angles, he must prepare
an unnatural Hamiltonian with exponentially-degenerated ground states.
This no-go theorem is general (we do not assume the state to be the
AKLT states), 
hence drastically new scheme of blind quantum computation
must be required in order to perform blind quantum computation
in the ground space of a natural Hamiltonian in the single-server
setting.
On the other hand, our double-server protocol is the first 
blind delegated quantum computing protocol in which Alice is completely 
classical and quantum computation is performed 
in the ground space of a physically-motivated Hamiltonian.

One might think that
why we have not considered
the direct generalization of the double-server
protocol in Ref.~\cite{blindcluster}.
Here, the direct generalization means following protocol: Alice sends
hidden angles $\{\xi_{a,b}^Z\}$ to Bob1, 
and Bob1 creates rotated AKLT states.
Bob1 then teleports his particles to Bob2, and Bob2 performs measurements.
Let us briefly explain why such a direct generalization does not work
for AKLT resource states.
If Bob1 teleports a prerotated particle by $\xi_{a,b}^Z$, 
the operation implemented in the correlation space by Bob2's measurement
can be unwanted $Z$ rotation $e^{i\xi_{a,b}^Z Z/2}$ 
for certain teleportation error and Bob2's measurement result.
In this case, Alice must compensate the accumulation of
these unwanted rotations, 
and therefore she must send some number which is related to $\{\xi_{a,b}^Z\}$.
Since Bob1 knows teleportation errors and
$\{\xi_{a,b}^Z\}$, he can gain some information
about Bob2's measurement results,
and therefore Bob2 can send some classical message to Bob1.

\begin{figure}[htbp]
\begin{center}
\includegraphics[width=0.5\textwidth]{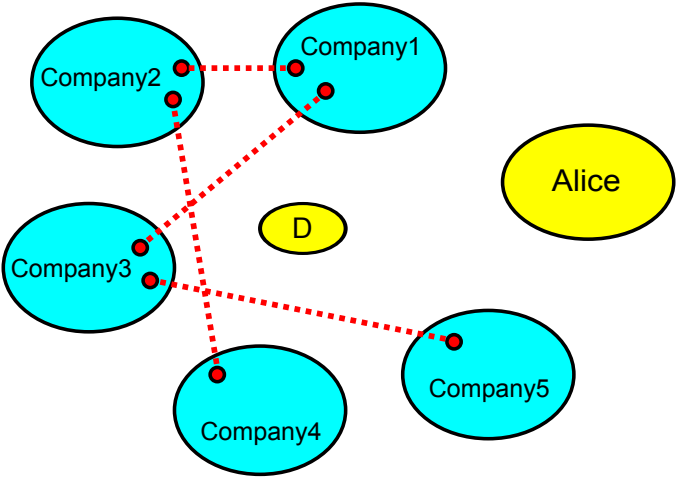}
\end{center}
\caption{(Color online.) 
D is the trusted third party which distributes Bell pairs.
Two red circles connected by a red dotted line are qubits of a single
Bell pair.
} 
\label{quantum_pepsi}
\end{figure}

Although the assumption in the double-server protocol
that two Bobs share no communication channel
seems to be somehow strong, it is not far from realistic.
For example,
let us consider an interesting future story where many competing companies 
offer services of quantum computation (Fig.~\ref{quantum_pepsi}).
A trusted third party (D) or the client (Alice) herself 
distributes many Bell pairs among these companies, 
but the information 
of which company shares a 
particular Bell pair with another particular company is kept secret
to all companies.
Alice then contacts two companies of her choice and runs the 
double-server protocol by instructing both servers which Bell pair 
components they have to use. 
Let us assume that Alice contacts Company1 and Company2,
where neither of them know with which company 
is doing the double-server protocol. 
Assume Company1 
wants to cheat Alice, hence 
Company1 
must call every other companies.
However, due to the competitive nature of the market, 
if Company1 contacts  
a wrong company, say Company3, 
Company3 gains the evidence that Company1  
is trying to cheat Alice, 
and Company3 can announce this fact
in order 
to destroy the credibility of Company1
thereby removing Company1 from the market.
Probably investing and maintaining a quantum server must be cost-expensive,
hence
no company wants to take the risk of being put out of the business.

At this stage, we do not know which quantum computation model (circuit,
adiabatic, measurement-based model, etc.) is the most promising candidate
for the realization of a scalable quantum computer.
Therefore, it is valuable to explore possibilities
of blind quantum computation in various models.
It would be an interesting subject of future study.

\acknowledgements
The authors would like to acknowledge the support from
Carnegie trust for the Universities of
Scotland and the 
Quantum Information Scotland Network (QUISCO)
for sponsoring TM visit which made this collaboration possible.
TM is supported by ANR (StatQuant JC07 07205763).
VD is supported by EPSRC (grant EP/G009821/1).
EK is supported by EPSRC (grant EP/E059600/1).

\appendix

\section{}
\label{App:Pinverse}
The inverse $P^\dagger$ of the PEPS $P$ can be implemented as follows.
Bob1 first prepares
the two-qubit state $|0\rangle\otimes|0\rangle$,
and performs the filtering operation
$\{F_{i,j}\}_{(i,j)=(0,0)}^{(1,1)}$
on the qutrit to which he wants to apply $P^\dagger$
and the two-qubit state,
where
\begin{eqnarray*}
F_{i,j}\equiv|1\rangle\langle1|\otimes |\eta_3\rangle\langle i,j|
+|2\rangle\langle2|\otimes |\eta_4\rangle\langle i,j|
+|3\rangle\langle3|\otimes |\eta_2\rangle\langle i,j|,
\end{eqnarray*}
$|i,j\rangle\equiv|i\rangle\otimes|j\rangle$, 
and
\begin{eqnarray*}
\sum_{i=0}^1
\sum_{j=0}^1
F_{i,j}^\dagger F_{i,j}=I.
\end{eqnarray*}
With probability 1, $F_{0,0}$ is realized.
Next Bob1 measures
the qutrit in the basis 
\begin{eqnarray*}
\Big\{
\frac{1}{\sqrt{3}}\Big(|1\rangle+|2\rangle+|3\rangle\Big),
\frac{1}{\sqrt{3}}
\Big(|1\rangle+e^{i2\pi/3}|2\rangle+e^{i4\pi/3}|3\rangle\Big),
\frac{1}{\sqrt{3}}
\Big(|1\rangle+e^{i4\pi/3}|2\rangle+e^{i8\pi/3}|3\rangle\Big)
\Big\}
\end{eqnarray*}
and corrects the phase error
if the result is the second or the third one.

\section{}
\label{App:tele2}
Let $E\otimes E'\in\{I,X,Z,XZ\}\otimes\{I,X,Z,XZ\}$ 
be the Pauli error on two qubits,
and define the PEPS $\tilde{P}$ by
\begin{eqnarray*}
\tilde{P}\equiv P(E\otimes E')P^\dagger P.
\end{eqnarray*}
Also define the matrices $\{\tilde{A}[1],\tilde{A}[2],\tilde{A}[3]\}$
by
\begin{eqnarray*}
\tilde{P}=\frac{1}{\sqrt{2}}\sum_{l=1}^3\sum_{a=0}^1\sum_{b=0}^1
\tilde{A}_{a,b}[l]|l\rangle\langle a,b|,
\end{eqnarray*}
where $\tilde{A}_{a,b}[l]$ depends on the Pauli error as summarized
in the second, third, and fourth columns of the following
table.
The fifth, sixth, and seventh columns are the summary of the operations
implemented in the correlation space by the measurement
${\mathcal M}(\phi)$ after the PEPS $\tilde{P}$ for each error.
The eighth, nineth, and tenth columns are the summary of the operations
implemented in the correlation space by the application of $V$ and the measurement
${\mathcal M}(\phi)$ after the PEPS $\tilde{P}$ for each error.
\begin{center}
\begin{tabular}{|c||c|c|c||c|c|c||c|c|c|}
\hline
$E\otimes E'$&$\tilde{A}[1]$&$\tilde{A}[2]$&$\tilde{A}[3]$&
$|\alpha(\phi)\rangle$&$|\beta(\phi)\rangle$&$|\gamma\rangle$&
$|\alpha(\phi)\rangle$&$|\beta(\phi)\rangle$&$|\gamma\rangle$\\
\hline
\hline
$I\otimes I$&
$X$&
$XZ$&
$Z$&
$Xe^{iZ\phi/2}$&
$XZe^{iZ\phi/2}$&
$Z$&
$XZe^{-iX\phi/2}$&
$Ze^{-iX\phi/2}$&
$X$\\
\hline
$I\otimes X$&
$0$&
$-Z$&
$-XZ$&
$-Z$&
$-Z$&
$-XZ$&
$-Ze^{-iX\phi/2}$&
$-XZe^{-iX\phi/2}$&
$0$\\
\hline
$I\otimes Z$&
$XZ$&
$X$&
$0$&
$XZe^{iZ\phi/2}$&
$Xe^{iZ\phi/2}$&
$0$&
$X$&
$X$&
$XZ$\\
\hline
$I\otimes XZ$&
$Z$&
$0$&
$-X$&
$Z$&
$Z$&
$-X$&
$-X$&
$-X$&
$Z$\\
\hline
$X\otimes I$&
$0$&
$Z$&
$XZ$&
$Z$&
$Z$&
$XZ$&
$Ze^{-iX\phi/2}$&
$XZe^{-iX\phi/2}$&
$0$\\
\hline
$X\otimes X$&
$X$&
$-XZ$&
$-Z$&
$Xe^{-iZ\phi/2}$&
$-XZe^{-iZ\phi/2}$&
$-Z$&
$-XZe^{-iX\phi/2}$&
$-Ze^{-iX\phi/2}$&
$X$\\
\hline
$X\otimes Z$&
$Z$&
$0$&
$X$&
$Z$&
$Z$&
$X$&
$X$&
$X$&
$Z$\\
\hline
$X\otimes XZ$&
$XZ$&
$-X$&
$0$&
$XZe^{-iZ\phi/2}$&
$-Xe^{-iZ\phi/2}$&
$0$&
$-X$&
$-X$&
$XZ$\\
\hline
$Z\otimes I$&
$-XZ$&
$-X$&
$0$&
$-XZe^{iZ\phi/2}$&
$-Xe^{iZ\phi/2}$&
$0$&
$-X$&
$-X$&
$-XZ$\\
\hline
$Z\otimes X$&
$Z$&
$0$&
$X$&
$Z$&
$Z$&
$X$&
$X$&
$X$&
$Z$\\
\hline
$Z\otimes Z$&
$-X$&
$-XZ$&
$Z$&
$-Xe^{iZ\phi/2}$&
$-XZe^{iZ\phi/2}$&
$Z$&
$-XZe^{iX\phi/2}$&
$Ze^{iX\phi/2}$&
$-X$\\
\hline
$Z\otimes XZ$&
$0$&
$-Z$&
$XZ$&
$-Z$&
$-Z$&
$XZ$&
$-Ze^{iX\phi/2}$&
$XZe^{iX\phi/2}$&
$0$\\
\hline
$XZ\otimes I$&
$Z$&
$0$&
$-X$&
$Z$&
$Z$&
$-X$&
$-X$&
$-X$&
$Z$\\
\hline
$XZ\otimes X$&
$-XZ$&
$X$&
$0$&
$-XZe^{-iZ\phi/2}$&
$Xe^{-iZ\phi/2}$&
$0$&
$X$&
$X$&
$-XZ$\\
\hline
$XZ\otimes Z$&
$0$&
$Z$&
$-XZ$&
$Z$&
$Z$&
$-XZ$&
$Ze^{iX\phi/2}$&
$-XZe^{iX\phi/2}$&
$0$\\
\hline
$XZ\otimes XZ$&
$-X$&
$XZ$&
$-Z$&
$-Xe^{-iZ\phi/2}$&
$XZe^{-iZ\phi/2}$&
$-Z$&
$XZe^{iX\phi/2}$&
$-Ze^{iX\phi/2}$&
$-X$\\
\hline
\end{tabular}
\end{center}

\section{}
\label{App:Xrot}

Protocol 6 describes how Alice performs the single-qubit
$x$-rotation
$\exp\Big[\frac{-iX}{2}\theta^X_{k,b}\Big]$ by 
$\theta^X_{k,b}\in{\mathcal A}$ 
in $(k,b)$th $X$-AKLT subsystem. 

\begin{algorithm}
\caption{\bf 6: Double-server Blind $X$ rotation}
\label{prot:2Xrot}

Initially the flag parameter (known to only Alice) is set $\tau=1$. 
Alice sets her secret parameter $\epsilon^X_{k,b}=0$ and also chooses 
random numbers $\delta^X_{k,b}\in{\mathcal A}$. 
Alice sends Bob1 parameter values $N$, $M$ and $n<N$. Bob1 creates $M$ AKLT chains $|AKLT^{N,L,R}_b\rangle$, where $b=1,...,M$ (see Equation \ref{eq:AKLT}) of $N$ qutrits arranged in an array of $N$ columns and $M$ rows. 
For $j=1, \cdots, n/2$, Alice, Bob1, and Bob2 repeat (I)-(VI), 
where the steps {\bf (I)},{\bf (II)}{\bf (III)} are the same as in Protocol 5.
\begin{itemize}
\item[\bf (IV)]
Alice sends the angle $\phi^X_{(k-1)n+j,b}$
to Bob2. This angle is determined according to the following rule:
\begin{itemize}

\item[$\bullet$]
If the Pauli error is 
$I\otimes I$,
$I\otimes X$,
$X\otimes I$,
or $X\otimes X$,
\begin{eqnarray*}
\phi^X_{(k-1)n+j,b}=\tau\theta^X_{k,b}+\tau\delta^X_{k,b}+\xi^X_{(k-1)n+j,b}+
r^X_{(k-1)n+j,b}\pi~~~(\mbox{mod}~2\pi),
\end{eqnarray*}
where $\xi^X_{(k-1)n+j,b}\in{\mathcal A}$ 
and $r^X_{(k-1)n+j,b}\in\{0,1\}$ are random numbers chosen by Alice,
signs of $\theta^X_{k,b}$ and $\delta^X_{k,b}$ should be changed if there is the
byproduct $Z$ before this step.

\item[$\bullet$]
If the Pauli error is 
$Z\otimes Z$,
$Z\otimes XZ$,
$XZ\otimes Z$,
or $XZ\otimes XZ$,
\begin{eqnarray*}
\phi^X_{(k-1)n+j,b}=-\tau\theta^X_{k,b}-\tau\delta^X_{k,b}-\xi^X_{(k-1)n+j,b}
+r^X_{(k-1)n+j,b}\pi~~~(\mbox{mod}~2\pi),
\end{eqnarray*}
where $\xi^X_{(k-1)n+j,b}\in{\mathcal A}$ 
and $r^X_{(k-1)n+j,b}\in\{0,1\}$ are random numbers chosen by Alice.
Signs of $\theta^X_{k,b}$ and $\delta^X_{k,b}$ should be changed if there is the
byproduct $Z$ before this step.

\item[$\bullet$]
If the Pauli error is
$I\otimes Z$,
$I\otimes XZ$,
$X\otimes Z$,
$X\otimes XZ$,
$Z\otimes I$,
$Z\otimes X$,
$XZ\otimes I$,
or
$XZ\otimes X$,
\begin{eqnarray*}
\phi^X_{(k-1)n+j,b}=\xi^X_{(k-1)n+j,b} 
\end{eqnarray*}
where
$\xi^X_{(k-1)n+j,b}\in{\mathcal A}$
is a random number chosen by Alice.
\end{itemize}

\item[\bf (V)]
Bob2 applies $V$, does the measurement $\mathcal{M}(\phi^X_{(k-1)n+j,b})$, 
and
sends the result to Alice.
By this measurement, following operations are implemented 
in the correlation space (see also Appendix~\ref{App:tele2}):
\begin{itemize}

\item[$\bullet$]
If the Pauli error is $I\otimes I$ or $X\otimes X$,
$|\alpha(\phi^X_{(k-1)n+j,b})\rangle$,
$|\beta(\phi^X_{(k-1)n+j,b})\rangle$,
or $|\gamma\rangle$ occurs with the probability $1/3$, respectively.
If 
$|\alpha(\phi^X_{(k-1)n+j,b})\rangle$ is realized,
\begin{eqnarray*}
X^{r^X_{(k-1)n+j,b}+1}Z
\exp\Big[\frac{-iX}{2}\Big(\tau\theta^X_{k,b}+\tau\delta^X_{k,b}+
\xi^X_{(k-1)n+j,b}\Big)\Big]
\end{eqnarray*}
is implemented.
If 
$|\beta(\phi^X_{(k-1)n+j,b})\rangle$ is realized,
\begin{eqnarray*}
X^{r^X_{(k-1)n+j,b}}Z\exp
\Big[\frac{-iX}{2}\Big(\tau\theta^X_{k,b}+\tau\delta^X_{k,b}+\xi^X_{(k-1)n+j,b}\Big)\Big]
\end{eqnarray*}
is implemented.
If $|\gamma\rangle$
is realized,
$X$ is implemented.
\item[$\bullet$]
If the Pauli error is $I\otimes X$ or $X\otimes I$,
$|\alpha(\phi^X_{(k-1)n+j,b})\rangle$ or
$|\beta(\phi^X_{(k-1)n+j,b})\rangle$
occurs with the probability $1/2$, respectively.
If 
$|\alpha(\phi^X_{(k-1)n+j,b})\rangle$ is realized,
\begin{eqnarray*}
X^{r^X_{(k-1)n+j,b}}Z
\exp\Big[\frac{-iX}{2}\Big(\tau\theta^X_{k,b}+\tau\delta^X_{k,b}+\xi^X_{(k-1)n+j,b}\Big)\Big]
\end{eqnarray*}
is implemented.
If 
$|\beta(\phi^X_{(k-1)n+j,b})\rangle$ is realized,
\begin{eqnarray*}
X^{r^X_{(k-1)n+j,b}+1}Z
\exp\Big[\frac{-iX}{2}\Big(\tau\theta^X_{k,b}+\tau\delta^X_{k,b}+\xi^X_{(k-1)n+j,b}\Big)\Big]
\end{eqnarray*}
is implemented.

\end{itemize}
\end{itemize}
\end{algorithm}

\begin{algorithm}
\caption{\bf 6: --- Continued}
\begin{itemize}
\item[\bf (V)] ---Continued
\begin{itemize}

\item[$\bullet$]
If the Pauli error is $Z\otimes Z$ or $XZ\otimes XZ$,
$|\alpha(\phi^X_{(k-1)n+j,b})\rangle$,
$|\beta(\phi^X_{(k-1)n+j,b})\rangle$,
or $|\gamma\rangle$ occurs with the probability $1/3$, respectively.
If 
$|\alpha(\phi^X_{(k-1)n+j,b})\rangle$ is realized,
\begin{eqnarray*}
X^{r^X_{(k-1)n+j,b}+1}Z
\exp\Big[\frac{-iX}{2}\Big(\tau\theta^X_{k,b}+\tau\delta^X_{k,b}+\xi^X_{(k-1)n+j,b}\Big)\Big]
\end{eqnarray*}
is implemented.
If 
$|\beta(\phi^X_{(k-1)n+j,b})\rangle$ is realized,
\begin{eqnarray*}
X^{r^X_{(k-1)n+j,b}}Z
\exp\Big[\frac{-iX}{2}\Big(\tau\theta^X_{k,b}+\tau\delta^X_{k,b}+\xi^X_{(k-1)n+j,b}\Big)\Big]
\end{eqnarray*}
is implemented.
If $|\gamma\rangle$ is realized,
$X$ is implemented.

\item[$\bullet$]
If the Pauli error is $Z\otimes XZ$ or $XZ\otimes Z$,
$|\alpha(\phi^X_{(k-1)n+j,b})\rangle$ or
$|\beta(\phi^X_{(k-1)n+j,b})\rangle$
occurs with the probability $1/2$, respectively.
If 
$|\alpha(\phi^X_{(k-1)n+j,b})\rangle$ is realized,
\begin{eqnarray*}
X^{r^X_{(k-1)n+j,b}}Z
\exp\Big[\frac{-iX}{2}\Big(\tau\theta^X_{k,b}+\tau\delta^X_{k,b}+\xi^X_{(k-1)n+j,b}\Big)\Big]
\end{eqnarray*}
is implemented.
If 
$|\beta(\phi^X_{(k-1)n+j,b})\rangle$ is realized,
\begin{eqnarray*}
X^{r^X_{(k-1)n+j,b}+1}Z
\exp\Big[\frac{-iX}{2}\Big(\tau\theta^X_{k,b}+\tau\delta^X_{k,b}+\xi^X_{(k-1)n+j,b}\Big)\Big]
\end{eqnarray*}
is implemented.

\item[$\bullet$]
If the Pauli error is $I\otimes Z$, $X\otimes XZ$, $Z\otimes I$,
or $XZ\otimes X$,
$|\alpha(\phi^X_{(k-1)n+j,b})\rangle$,  
$|\beta(\phi^X_{(k-1)n+j,b})\rangle$,
or $|\gamma\rangle$ occurs with the probability
$\frac{1}{2}\cos^2[\frac{1}{2}\phi^X_{(k-1)n+j,b}]$,
$\frac{1}{2}\sin^2[\frac{1}{2}\phi^X_{(k-1)n+j,b}]$,
or $1/2$, respectively.
If 
$|\alpha(\phi^X_{(k-1)n+j,b})\rangle$
or
$|\beta(\phi^X_{(k-1)n+j,b})\rangle$
is realized, $X$ is implemented.
If $|\gamma\rangle$ is realized,
$XZ$ 
is implemented.

\item[$\bullet$]
If the Pauli error is $I\otimes XZ$, $X\otimes Z$, $Z\otimes X$,
or $XZ\otimes I$,
$|\alpha(\phi^X_{(k-1)n+j,b})\rangle$,  
$|\beta(\phi^X_{(k-1)n+j,b})\rangle$,
or $|\gamma\rangle$ occurs with the probability
$\frac{1}{2}\sin^2[\frac{1}{2}\phi^X_{(k-1)n+j,b}]$,
$\frac{1}{2}\cos^2[\frac{1}{2}\phi^X_{(k-1)n+j,b}]$,
or $1/2$, respectively.
If 
$|\alpha(\phi^X_{(k-1)n+j,b})\rangle$
or
$|\beta(\phi^X_{(k-1)n+j,b})\rangle$
is realized, $X$ is implemented.
If $|\gamma\rangle$ is realized,
$Z$ 
is implemented.

\end{itemize}

\item[\bf (VI)]
If the $x$-rotation 
$\exp[\frac{-iX}{2}(\tau\theta^X_{k,b}+\tau\delta^X_{k,b}+\xi^X_{(k-1)n+j,b})]$
is implemented in the previous step, Alice sets $\tau=0$, 
and
$\epsilon^X_{k,b}=\epsilon^X_{k,b}+
\xi^X_{(k-1)n+j,b}$ $(\mbox{mod}~2\pi)$ (if there is no $Z$ byproduct before
this rotation) or
$\epsilon^X_{k,b}=\epsilon^X_{k,b}-
\xi^X_{(k-1)n+j,b}$ $(\mbox{mod}~2\pi)$ (if there is the $Z$ byproduct before
this rotation).

\item[\bf(VII)]
So far, 
the $x$-rotation 
\begin{eqnarray*}
G^X_{k,b}\exp\Big[\frac{-iX}{2}
(\epsilon^X_{k,b}+\delta^X_{k,b})\Big]
\exp\Big[\frac{-iX}{2}\theta^X_{k,b}\Big]
\end{eqnarray*}
up to a Pauli byproduct $G^X_{k,b}$ is implemented.
The probability that they fail to perform this rotation
is $(2/3)^{n/2}$,
which is small for sufficiently large $n$.
Alice asks Bob1 to correct the accumulated error. In order to do so she sends Bob1 the angle 
$\tilde{\epsilon}^X_{k,b}=\epsilon^X_{k,b}+\delta^X_{k,b}$ 
$(\mbox{mod}~2\pi)$ if $G^X_{k,b}$ contains
no $Z$ byproduct, and 
$\tilde{\epsilon}^X_{k,b}=-\epsilon^X_{k,b}-\delta^X_{k,b}$ 
$(\mbox{mod}~2\pi)$ if
$G^Z_{k,b}$ contains the $Z$ byproduct.
Bob1 implements the rotation
$\exp\Big[\frac{iX}{2}\tilde{\epsilon}^X_{k,b}\Big]$
by using the rest of the qutrits in $(k,b)$th AKLT subsystem.
The probability that Bob1 fails to perform this $x$-rotation
is $(1/3)^{n/2}$,
which is small for sufficiently large $n$.

\end{itemize}
\end{algorithm}


\begin{thebibliography}{00}

\bibitem{Childs}
A. Childs, Quant. Inf. Compt. {\bf5}, 456 (2005).

\bibitem{Arrighi}
P. Arrighi and L. Salvail, Int. J. Quant. Inf. {\bf4}, 883 (2006).

\bibitem{blindcluster}
A. Broadbent, J. Fitzsimons, and E. Kashefi,
Proceedings of the 50th Annual IEEE Symposium on Foundations of
Computer Science (FOCS 2009), pp.517.

\bibitem{Aha}
D. Aharonov, M. Ben-Or, and E. Eban,
Proceedings of Innovations in Computer Science 2010 (ICS 2010), pp. 453.

\bibitem{Shor}
P. Shor, Proceedings of the 35th Symposium on Foundations of Computer Science,
Los Alamitos, edited by Shafi Goldwasser (IEEE Computer
Society Press), pp. 124.

\bibitem{Feigenbaum}
J. Feigenbaum, Proceedings of Advances in Cryptology, CRYPTO 85, 1986, pp.477.

\bibitem{NP}
Non-deterministic polynomial-time hard function is a problem that is at least 
as hard as the hardest problems in the class {\bf NP}, where {\bf NP} is the class of all problems 
whose solutions are easy to verify, whereas finding a solution is considered intractable.


\bibitem{Abadi}
M.~Abadi, J.~Feigenbaum, and J.~Kilian,
Journal of Computer and System Sciences,1989, {\bf 39}, pp. 21.


\bibitem{Gentry}
C. Gentry, 
Proceedings of the 41st annual ACM Symposium on Theory of Computing (STOC 2009), pp. 169.

\bibitem{onetimepad}
P. Boykin and V. Roychowdhury, Phys. Rev. A {\bf67}, 042317 (2003).

\bibitem{cluster}
R. Raussendorf and H. J. Briegel, Phys. Rev. Lett. {\bf86}, 5188 (2001).
\bibitem{cluster2}
R. Raussendorf, D. E. Browne, and H. J. Briegel, 
Phys. Rev. A {\bf68}, 022312 (2003).

\bibitem{VerstraeteVBS}
F. Verstraete and J. I. Cirac,
Phys. Rev. A {\bf70}, 060302(R) (2004).

\bibitem{MiyakeAKLT}
G. K. Brennen and A. Miyake, Phys. Rev. Lett. {\bf101}, 010502 (2008).

\bibitem{Gross}
D. Gross and J. Eisert, Phys. Rev. Lett. {\bf98}, 220503 (2007).
\bibitem{Gross2}
D. Gross, J. Eisert, N. Schuch, and D. Perez-Garcia,
Phys. Rev. A {\bf76}, 052315 (2007).
\bibitem{Gross3}
D. Gross and J. Eisert, 
Phys. Rev. A {\bf82}, 040303(R) (2010).

\bibitem{Miyakeholographic}
A. Miyake, Phys. Rev. Lett. {\bf105}, 040501 (2010).

\bibitem{Miyake2dAKLT}
A. Miyake, arXiv: 1009.3491.

\bibitem{Wei}
T. C. Wei, I. Affleck, and R. Raussendorf,
Phys. Rev. Lett. {\bf106}, 070501 (2011).


\bibitem{Caimagnet}
J. Cai, A. Miyake, W. D\"ur, and H. J. Briegel,
Phys. Rev. A {\bf82}, 052309 (2010).

\bibitem{Li}
Y. Li, D. E. Browne, L. C. Kwek, R. Raussendorf, and T. C. Wei,
arXiv: 1102.5153.

\bibitem{tricluster}
X. Chen, B. Zeng, Z. Gu, B. Yoshida, and I. L. Chuang,
Phys. Rev. Lett. {\bf102}, 220501 (2009).
\bibitem{reduction}
X. Chen, R. Duan, Z. Ji, and B. Zeng,
Phys. Rev. Lett. {\bf105}, 020502 (2010).

\bibitem{AKLT}
I. Affleck, T. Kennedy, E. H. Lieb, and H. Tasaki,
Comm. Math. Phys. {\bf115}, 477 (1988).

\bibitem{Haldane}
F. D. M. Haldane,
Phys. Rev. Lett. {\bf50}, 1153 (1983).

\bibitem{string}
M. den Nijs and K. Rommelse, Phys. Rev. B {\bf40}, 4709 (1989).

\bibitem{edge}
T. Kennedy, J. Phys. Condens. Matter {\bf2}, 5737 (1990).

\bibitem{Darmawan}
A. S. Darmawan and S. D. Bartlett,
Phys. Rev. A {\bf82}, 012328 (2010).

\bibitem{Garcia}
D. Perez-Garcia, F. Verstraete, M. M. Wolf, and J. I. Cirac,
Quant. Inf. Comput. {\bf7}, 401 (2007).
\bibitem{Verstraete_review}
F. Verstraete, J. I. Cirac, and V. Murg,
Adv. Phys. {\bf57}, 143 (2008).
\bibitem{Cirac_review}
J. I. Cirac and F. Verstraete, J. Phys. A: Math. Theor. {\bf42}, 504004 (2009).
\bibitem{localizable}
F. Verstraete, M. Popp, and J. I. Cirac, Phys. Rev. Lett. {\bf92},
027901 (2004).

\bibitem{Delgado}
F. Verstraete, M. A. Martin-Delgado, and J. I. Cirac,
Phys. Rev. Lett. {\bf92}, 087201 (2004).
\bibitem{Bobcandofiltering}
Here we assume that Alice performs the projection
$I-|\eta_1\rangle\langle\eta_1|$,
since it makes our explanations simpler.
Actually, this projection can be done by Bob.

\bibitem{pi8_universal}
P. O. Boykin, T. Mor, M. Pulver, V. Roychowdhury, and F. Vatan,
Proceedings of 40th Annual Symposium on Fondations of Computer 
Science (FOCS)
(IEEE Press, Los Alamitos, CA, 1999), pp.486-494;
arXiv:9906054.



\bibitem{uploading}
T. Morimae, Phys. Rev. A {\bf83}, 042337 (2011).

\bibitem{inBob'sbrain}
Physically, PEPS operation just means Bob encodes the three-dimensional subspace
of the four-dimensional Hilbert space as a single qutrit.

\end{thebibliography}
\end{document}